\documentclass[11pt]{article}
\usepackage{amsmath, braket, amsthm, amssymb, enumerate, graphicx, subcaption, longtable, array} 
\usepackage[margin = 2cm]{geometry}
\usepackage[colorlinks=true]{hyperref}

\DeclareMathOperator{\nbd}{nbd}
\DeclareMathOperator{\Nbd}{Nbd}

\DeclareMathOperator{\trace}{trace}

\newtheorem{definition}{Definition}
\newtheorem{procedure}{Procedure}
\newtheorem{assumption}{Assumption}

\title{Discrete-Time Open Quantum Walks for Vertex Ranking in Graphs}
\author{Supriyo Dutta\\
\small{Department of Mathematics}, \\ \small{National Instititute of Technology Agartala,} \\ \small{Jirania,  West Tripurs, India - 799046.} \\ \small{Email: \texttt{dosupriyo@gmail.com}}}
\date{} 

\begin{document}

	\maketitle 

	\begin{abstract}
		\noindent This article presents a new quantum PageRank algorithm on graphs using discrete-time open quantum walks. Google's PageRank is a widely used algorithm for ranking the web pages on the World Wide Web in classical computation. From a broader perspective, it is also a fundamental measure for quantifying the importance of vertices in a network. Similarly, the new quantum PageRank also serves to quantify the significance of a network's vertices. In this work, we extend the concept of discrete-time open quantum walk on arbitrary directed and undirected graphs by utilizing the Weyl operators as Kraus operators. This new model of quantum walk is useful for building up the quantum PageRank algorithm, discussed in this article. We compare the classical PageRank and the newly defined quantum PageRank for different types of complex networks, such as the scale-free network, Erd\H{o}s-R\'enyi random network, Watts-Strogatz network, spatial network, Zachary Karate club network, GNC, GN, GNR networks, Barabási and Albert network, etc. In addition, we study the convergence of the quantum PageRank process and its dependency on the damping factor $\alpha$. We observe that this quantum PageRank procedure is faster than many other proposals reported in the literature.
	\end{abstract}
	
	\tableofcontents
	
	\section{Introduction} 
		
		The presence of networks in our modern life is ubiquitous \cite{estrada2012structure}. As a consequence, there has been significant research in the field of network analysis, which is applied to the World Wide Web (WWW) \cite{gillies2000web}, and various types of virtual, social, economic, and biological systems \cite{strogatz2001exploring, boccaletti2006complex}. Collecting useful information from a network is a crucial task in network science. In classical computing, there are efficient algorithms for searching, retrieving and ordering information which is available from the network. In quantum computing, building a quantum network is a fundamental goal in quantum communication \cite{fang2023quantum} as it might be useful in developing the future internet and a scalable quantum computer. Thus, we need quantum mechanical algorithms which can efficiently interact with the information available in a quantum network \cite{nokkala2024complex}.  
		
		In classical computing, Google's PageRank algorithm \cite{page1999pagerank} is well-known for arranging the information available in WWW \cite{langville2006google}. The WWW is a collection of pages or documents with information. It has a crucial hyperlink structure. A hyperlink is a text on a page that carries the reader to another page. Mathematically, WWW is a directed graph whose vertices are the pages and the directed edges are the hyperlinks from one page to the other. The PageRank algorithm arranges the pages in WWW based on their importance, such that, a reader can go through the most relevant page, first. In general, it indicates the importance of the vertices in a graph. The PageRank algorithm is a polynomial -time algorithm from the perspective of computational complexity, which utilizes the idea of the random walk on a directed graph. We include a brief description of PageRank in section \ref{PageRank}.
		
		There are multiple attempts to develop the PageRank algorithm in the literature of quantum computing which we can precisely classify into two different approaches. In one of the initial works on \textbf{quantum PageRank (qPageRank)}, an adiabatic quantum computing algorithm was demonstrated \cite{garnerone2012adiabatic}. The other proposals replace the random walk of classical PageRank with a quantum walk. In quantum computing we quantize the idea of random walk on the graphs with quantum walk \cite{portugal2013quantum}. There are different proposals of quantum walk, such as the discrete-time quantum walk \cite{aharonov2001quantum, szegedy2004quantum, szegedy2004spectra}, continuous-time quantum walk \cite{farhi1998quantum, godsil2008periodic}, open quantum walk \cite{attal2012open}, and many others. A number of articles that use continuous-time quantum walk for developing qPageRank are \cite{boito2023ranking, boito2021quantum, wang2022continuous, izaac2017centrality}. The qPageRank based on the discrete-time quantum walk is proposed in \cite{paparo2012google}. There are articles which quantize the notion of centrality using quantum walk for instance \cite{boito2021quantum, wang2022continuous}. In \cite{sanchez2012quantum} qPageRanks based on continuous-time open quantum walk are discussed. Articles \cite{paparo2013quantum, paparo2014quantum, chawla2020discrete}, and \cite{ortega2023generalized} study different aspects of qPageRank based on Szegedy's quantum walk. Article \cite{loke2017comparing} compares the classical and quantum PageRanks on different types of networks. A number of classical computing aspects of qPageRank are discussed in \cite{tang2021tensorflow}. Article \cite{chapuis2023new} utilizes graph Fourier transformation on directed graphs and the quantum HHL algorithm \cite{harrow2009quantum} to calculate the classical PageRank. A qPageRank based on the stochastic quantum walk is proposed in \cite{benjamin2024resolving}. All these articles include numerical experiments on various models of complex networks to establish the justification for the efficiency of the newly developed algorithms, such as the outerplanar hierarchical network \cite{paparo2013quantum, loke2017comparing, chapuis2023new}, the scale-free network \cite{paparo2013quantum, loke2017comparing, chawla2020discrete, ortega2023generalized, chapuis2023new, benjamin2024resolving}, Erdős–Rényi random network \cite{loke2017comparing, chapuis2023new, benjamin2024resolving}, Watts-Strogatz network \cite{benjamin2024resolving}, spatial network \cite{benjamin2024resolving}, Zachary Karate club network \cite{benjamin2024resolving}, random\_k\_out\_graph \cite{boito2023ranking}, binary tree graph \cite{chawla2020discrete}, GNC network \cite{chawla2020discrete}, Barabási and Albert network \cite{chapuis2023new}, etc. The PageRank is an important measure of centrality on graphs. The experimental realization of qPageRank and centrality measures are discussed in \cite{wang2020experimental, izaac2017centrality, wu2020experimental}. Although there are many proposals of quantum PageRank in literature, none of them exactly match with the classical PageRank for all graphs due to the difference between classical and quantum random walk models. As qPageRank is not unique, we assume that they should satisfy the following assumptions which were proposed in \cite{paparo2012google}:
		\begin{assumption}\label{qPageRank_assumptions}
			A qPageRank should satisfy the following assumptions: 
			\begin{enumerate}[(i)]
				\itemsep0em 
				\item 
					The classical PageRank is defined on directed graphs. Hence, the qPageRank should also be defined on directed graphs.
				\item 
					The values of qPageRank of the vertices should be in $[0, 1]$. Also, the sum of all qPageRanks in a graph is $1$.
				\item 
					The qPageRank must admit a quantized Markov Chain description.
				\item 
					The classical algorithm to compute qPageRank of the vertices in a graph belongs to the computational complexity class \cite{arora2009computational} BQP. This criterion can be further restricted to enforce that the computational complexity of qPageRank must be $P$.
			\end{enumerate}
		\end{assumption} 
		
		The classical algorithm for PageRank is an eigenvalue finding problem and does not consider quantum interference at all. Therefore, it is essential to develop a quantum mechanical algorithm for ranking the vertices in a network. The literature crystallizes the crucial role of various quantum walks in developing the concept of qPageRank. Most of the quantum walks are originally defined on undirected graphs. To break this limitation, we introduce the open quantum walk, which is defined on directed graphs. Recall that the WWW is itself a directed graph, and the original idea of PageRank is defined on directed graphs. Since we aim to define qPageRank for all graphs, it is crucial to properly define the quantum walk for all graphs, particularly for directed graphs. The idea of an open quantum walk adds another dimension. Unlike the other quantum walks, the dynamics of the open quantum walk is non-unitary. It allows us to incorporate Markovian dynamics. It assures convergence of the instantaneous qPagerank sequence. For the other qPageRank algorithms, we require a sequence of time averages for convergence, but our particular choice of quantum walk eliminates this requirement. Our article uses the discrete-time open quantum walk, which was not employed before to quantify PageRank because it was mostly studied on a few undirected regular graphs, like the infinite path graph. or $\mathbb{Z}$, cycle \cite{attal2012open, sinayskiy2013open, dhahri2019open}. To construct qPageRank with this quantum walk, we extend its definition to all graphs. This quantum walk model primarily employs the Kraus operators . We observe that the Weyl operators \cite{bennett1993teleporting, bertlmann2008bloch} can also be utilized to construct the Kraus operators \cite{basile2024weyl}. We extend the idea of an open quantum walk for arbitrary directed and undirected graphs. To the best of our knowledge, we employ the Weyl operators to define the open quantum walk for the first time in literature. 
		
		Our fundamental contribution to this article is the proposal of a novel qPageRank using the discrete-time open quantum walk. The discrete-time open quantum walk utilizes a Hilbert space of dimension $n^2$, where $n$ is the number of vertices in the graph. We demonstrate that it is possible to reduce the size of the Hilbert space to a subspace of dimension $n$ for the purpose of computing the qPageRank on a graph, a task that has proven to be challenging in the literature on the subject. We calculate the qPageRank for the vertices of a number of directed and undirected networks, such as the scale-free network, Erd\H{o}s–R\'enyi random network, Watts-Strogatz network, Zachary Karate club network, random\_k\_out\_graph, tree graph, GNC network, Barabási and Albert network, etc. We justify that the new qPageRank algorithm also satisfies all the required assumptions that should be fulfilled for a qPageRank, mentioned in Assumption \ref{qPageRank_assumptions}. Similar to other qPageRank models, the new qPageRank may not always align with the classical PageRank for all graph types. The similarity and dissimilarity between PageRank and qPageRank depend on the graphs. We also check the convergence of the qPageRank process. We observe that our PageRank procedure converges faster than many other qPageRank procedures reported in literature \cite{loke2017comparing, chawla2020discrete, ortega2023generalized}. In addition, we check the dependency of qPageRank on the damping factor $\alpha$ for different graphs.
		
		This article is organized as follows: Section 2 compiles all the preliminary ideas for this article. It has three subsections. We begin with an introduction to graph theory, which is the mathematical foundation of the network analysis. In section \ref{PageRank}, we discuss the original idea of classical PageRank. We end this section with a discussion on quantum states and channels in higher dimensions. In Section 3, we introduce a new concept by demonstrating a generalization of the discrete-times open quantum walk for any directed graphs. Section 4 models how an internet surfer can restart surfing from another randomly chosen node in a quantum mechanical way, which is a crucial assumption in the classical PageRank algorithm. The qPageRank of a graph's vertices is the focus of Section 5. We discuss our qPageRank algorithm in Procedure \ref{qPageRank_algo}. In this section, we justify that a classical computer can calculate the new PageRank algorithm in polynomial time on an $n$-dimensional Hilbert space. We divide our numerical experiments into four subsections in Section 6. In the first two subsections, we compare the classical and quantum PageRanks for a number of undirected and directed graphs, respectively. We dedicate the third subsection to the convergence of the qPageRank procedure. The last subsection presents the dependency of qPageRank on the primitivity parameter. Then, we conclude this article.

	\section{Preliminaries} 
	
		\subsection{An introduction to graph theory}
		
			A graph $G = (V(G), E(G))$ is a combinatorial object consisting of a set of vertices $V(G)$ and a set of edges $E(G) \subset V(G) \times V(G)$ \cite{west2001introduction}. Throughout this article, $n$ denotes the number of vertices in a graph $G$. An edge $(u, v)$ between the vertices $u$ and $v$ is said to be incident to the vertices $u$ and $v$. Also, $u$ and $v$ are said to be the neighbors of each other. We employ different types of edges in this article. We denote a directed edge or link from the vertex $u$ to a vertex $v$ by $\overrightarrow{(u, v)}$. An undirected edge between vertices $u$ and $v$ is denoted by $(u, v)$, which can be considered as a combination of two directed edges, one from $u$ to $v$, that is $\overrightarrow{(u, v)}$ and another from $v$ to $u$, that is $\overrightarrow{(v, u)}$. In this article, the graphs have no loop on the vertices, which is an edge to join a vertex with itself. Also, there are no weights on the vertices and on the edges. All the edges of a directed graph are directed. Also, all the edges of an undirected graph are undirected, throughout this article.
			
			We pictorially represent a vertex with a dot which may be marked or non-marked. An undirected edge between vertices $u$ and $v$ is depicted by a straight line between $u$ and $v$. A directed edge from the vertex $u$ to the vertex $v$ is represented by an arrow from $u$ to $v$. We denote a directed graph by $\overrightarrow{G}$ and its edge set by $E(\overrightarrow{G})$.
			
			In a directed graph $\overrightarrow{G}$, the directed edges or links incident to a vertex $v$ may be classified into inlinks and outlinks. The directed edges $\overrightarrow{(u, v)}$ coming into a vertex $v$ from a vertex $u$ are called inlinks of $v$. The number of inlinks of $v$ is called the indegree of $v$ which is denoted by $d^{(i)}_v$, where $(i)$ indicate inlink. The set of vertices $u$ for which there is an inlink $\overrightarrow{(u, v)}$ to $v$ is denoted by $B(v)$, in this article. If there is a link from $u$ to $v$, we say that the vertex $u$ is pointing to $v$. Similarly, the outlinks of vertex $v$ are the directed edges $\overrightarrow{(v, u)}$ going out of the vertex $v$ to another vertex $u$. The number of outlinks of $v$ is its outdegree, which is denoted by $d^{(o)}_v$, where $(o)$ indicate outlink. Also, the set of vertices $u$ for which there is an outlink $\overrightarrow{(v, u)}$ from $v$ is denoted by $\Nbd(v)$, in this article. It is not necessary that all the vertices are incident to both types of edges. A dangling vertex in a graph has no outlink. 
			
			In an undirected graph $G$, a vertex $v$ is said to be adjacent to another vertex $u$ if there is an edge $(u, v) \in E(G)$. The degree of a vertex $v$ is denoted by $d_v$ which is the number of vertices $u$ adjacent to $v$. The set of the adjacent vertices of $v$ is called the neighbourhood of $v$ and denoted by $\nbd(v)$. Every undirected graph can be considered as a directed graph by assigning two opposite directions on every edge. Therefore, in an undirected graph, we have $\nbd(v) = B(v) = \Nbd(v)$ and $d_v = d^{(o)}_v = d^{(i)}_v$, for every vertex $v$. Also, there is no dangling vertex in an undirected graph.
			
			For our numerical experiments we use a number of graphs and networks, which may be directed or undirected. For instance, we use the path graph, cycle graph, complete graph, star graph, balanced tree graph, wheel graph, etc. Also we employ the scale-free network, Erd\H{o}s–R\'enyi random network, Watts-Strogatz network, Zachary Karate club network, random\_k\_out\_graph, tree graph, GNC, GN, GNR network, Barabási and Albert network, etc. We avoid individual descriptions of them to make the article shorter. The reader may go through the documentation of NetworkX for further details \cite{hagberg2008exploring}.

		\subsection{An introduction to Google's PageRank}\label{PageRank}
			
			To build up the mathematical model of information retrieval from WWW, we utilize the idea of an internet graph, which is a directed graph. Recall that the vertices of the internet graph represent webpages in WWW. The directed edges from one vertex to another represent the hyperlinks from one page to another, correspondingly.
			\begin{definition}
				The PageRank of a vertex $v$ in a directed graph $\overrightarrow{G}$ is denoted $r(v)$ and defined by
				\begin{equation}\label{rank_definition}
					r(v) = \sum_{u \in B(v)} \frac{r(u)}{d^{o}_u},
				\end{equation}
				where $B(v)$ is the set of vertices pointing into $v$ and $d^{o}_u$ is the number of outlinks from page $u$.
			\end{definition}
			As equation (\ref{rank_definition}) needs the PageRanks of all adjacent vertices of $v$ to calculate its PageRank, the inventors defined an iterative process. Let $r_k(v)$ be the PageRank of page $v$ at iteration $k$, then
			\begin{equation}\label{iterated_PageRank_definition}
				r_{k + 1}(v) = \sum_{u \in B(v)} \frac{r_k(u)}{d^{o}_u}.
			\end{equation}
			Initially, $r_0(v) = \frac{1}{n}$ for all vertices $v$, where $n$ is the total number of vertices in $\overrightarrow{G}$. Let $\Pi^{(k)}$ be a vector of order $1 \times n$ which contains the PageRanks of the vertices at $k$-th iteration for $k = 0, 1, 2, \dots$. Initially, we consider that the walker starts walking from any vertex of the graph, which is chosen uniformly and randomly. Therefore, the initial PageRank vector is $\Pi^{(0)} = \frac{1}{n}(1, 1, \dots 1(n$-times)). Here, the reader may notice that $\Pi^{(k)}$ represents a probability distribution for all $k$. The hyperlink matrix of the graph is defined by $H = (h_{u,v})_{n \times n}$, where 
			\begin{equation}
				h_{u,v} = \begin{cases} \frac{1}{d^{o}_u} & ~\text{if}~ \overrightarrow{(u, v)} \in E(\overrightarrow{G}); \\ 0 & ~\text{otherwise}. \end{cases}
			\end{equation}
			Equation (\ref{iterated_PageRank_definition}) suggests that in the $(k + 1)$-th iteration the PageRank vector is updated to 
			\begin{equation}
				\Pi^{(k + 1)} = \Pi^{(k)} H,
			\end{equation}
			for $k = 0, 1, 2, \dots$. To ensure convergence of this iterative process we make the following adjustments to the matrix $H$:
			\begin{enumerate}
				\itemsep0em 
				\item 
					\textbf{Stochasticity adjustment:} Let $v$ be a dangling node. Recall that there are no outgoing edges from $v$ to any other vertex. After entering a dangling node, we assume that the walker can move to any page randomly. Therefore, we replace the matrix $H$ by $S = H + \frac{1}{n} a e^\dagger$, where $a$ and $e$ are the column vectors of order $n$. The vector $e$ is the all $1$ vector. The vector $a$ consists of the elements $a_v = 1$ if $v$ is a dangling node and $0$ otherwise.
				\item 
					\textbf{Primitivity adjustment and restart in surfing:} Although the surfer follows the hyperlink structure of the Web, sometimes they may get bored. Then, they abandon the hyperlink method of surfing by teleporting to a new destination, where he begins hyperlink surfing again until the next teleportation. Considering $(1 - \alpha)$ as the probability of teleportation, we can define Google matrix as 
					\begin{equation}\label{Google_matrix}
						\mathcal{G} = \alpha S + \frac{(1 - \alpha)}{n} e e^\dagger.
					\end{equation} 
					Here, $\alpha$ is called the damping factor. The value $\alpha = 0.85$ is widely accepted.
			\end{enumerate}
			
			After all the above adjustments, we define an iterative  power method finding PageRank which is
			\begin{equation}
				\Pi^{(k + 1)} = \Pi^{(k)} \mathcal{G}.
			\end{equation}
			It can be proved that $\mathcal{G}$ ensures the convergence $\{\Pi^{(k)}\}$. Therefore, $\lim_{k \rightarrow \infty} \Pi^{(k)} = \Pi$ exists. We call $\Pi$ as a PageRank vector which is unique for a graph.

		\subsection{Quantum states and channels in higher dimensional system}
		
			In quantum mechanics and information theory, a pure state is equivalent to a normalized column vector in a Hilbert space $\mathcal{H}^n$ of dimension $n$. In this article, we utilize Dirac's bra-ket notation, only  to denote the state vectors. Therefore, $\ket{\psi}$ denotes a pure state and $\bra{\psi}$ is its conjugate transpose. There are a number of vectors in this article, such as the PageRank vector, which may not satisfy the condition of normalization. We do not use the bra-ket notation for them. We consider the standard basis of $\mathcal{H}^n$ which is $\{\ket{i}: \ket{i} = (0, 0, \dots 1 (i\text{-th position}), \dots 0)^\dagger\}$. A mixed quantum state in $\mathcal{H}^n$ is represented by a density matrix $\rho$, which is a positive semidefinite Hermitian matrix of order $n$ and $\trace(\rho) = 1$.
			
			A quantum channel is a communication channel that can transmit quantum information. Mathematically, the quantum channels are completely positive, trace-preserving maps between the spaces of operators. This work is developed on finite-dimensional spaces. Here, a quantum channel $\Psi: C^{n \times n} \rightarrow C^{n \times n}$ can be represented by a set of Kraus operators $\mathcal{K} = \{K_i:i = 1, 2, \dots d; d \leq n\}$, such that $\sum_{i = 1}^d K_i^\dagger K_i = I_n$ \cite{sudarshan1961stochastic, kraus1983states}. When we transmit a quantum state represented by a density matrix $\rho$ of order $n$ via a quantum channel $\Psi$ represented by $\mathcal{K}$ we find another density matrix 
			\begin{equation}
				\Psi(\rho) = \sum_{i = 1}^d K_i \rho K_i^\dagger.
			\end{equation}
			In this work, we construct the Kraus operator using the Weyl operators \cite{bennett1993teleporting, bertlmann2008bloch, dutta2023qudit}, which are unitary operators. In general, they are defined by 
			\begin{equation}\label{Weyl_operator}
				U_{r,s} = \sum_{i = 0}^{n - 1} \exp\left(\frac{2\pi \iota}{n}\right)^{ir} \ket{i} \bra{i \oplus s} ~\text{for}~ 0 \leq r, s \leq n - 1,
			\end{equation}
			where $\oplus$ denotes addition modulo $n$. Note that $U_{r,s}$ are unitary operators. Thus, we have $U_{r,s}^\dagger U_{r,s} = I_n$ for all $r, s$.

	\section{Discrete-time open quantum walk on arbitrary directed graph}
	\label{section_3}
	
		In this section, we define discrete time open quantum walk for arbitrary directed graphs. As we can convert an undirected graph to a directed graph, our construction is applicable for the undirected graphs, also. Let us assume that at time $t$ the walker is present at the vertex $v$, that has an out-degree $d_v^{(o)} \neq 0$. There are directed edges from the vertex $v$ to each of $u_i$ for $i = 1, 2, \dots d_v^{(o)}$, anywhere the walker can move at time $(t + 1)$. Also, we consider that the walker may not move from vertex $v$ at time $(t + 1)$. Therefore, the set of vertices where the walker may be available at time $(t + 1)$ is $\Nbd(v) \cup \{v\} = \{v, u_1, u_2, \dots u_d\}$. If for a vertex $v$ the out-degree $d_v^{(o)} = 0$, then $\Nbd(v) \cup \{v\} = \{v\}$. In this case the walker can not move from the vertex $v$.
		
		The movement of a walker of discrete-time quantum walk is determined by a set of coin operators and shift operators. Now, we construct a set of coin operators for every vertex $v$. Note that in a general graph degree of different vertices may be distinct. Let $(v, u_1), (v, u_2), \dots (v, u_{d_v^{(o)}})$ be the outgoing edges from vertex $v$. Now, we construct $(d_v^{(o)} + 1)$ Weyl operators $I_n, U_{v, u_1}, U_{v, u_2}, \dots U_{v, u_{d_v^{(o)}}}$ where $I_n$ acts for no movement from vertex $v$ and $U_{v, u_i}$ acts for the movement from vertex $v$ to $u_i$. Also, if the vertex $v$ is a pendent vertex, then the walker can not move to any other vertex from $v$ using the hyperlink structure of the internet graph. Then, only $I_n$ will be applied to lead the quantum dynamics. As $U_{v, u_i}$ are unitary operators for $i = 1, 2, \dots d_v^{(o)}$, we have
		\begin{equation}\label{Kraus_condition_of_Weyl_operator_combination}
			\begin{split}
				& I_n I_n^\dagger + U_{v, u_1}^\dagger U_{v, u_1} + U_{v, u_2}^\dagger U_{v, u_2} + \dots + U_{v, u_{d_v^{(o)}}}^\dagger U_{v, u_{d_v^{(o)}}} = (d_v^{(o)} + 1)I_n \\
				\text{or}~ & \frac{1}{\sqrt{(d_v^{(o)} + 1)}}I_n \frac{1}{\sqrt{(d_v^{(o)} + 1)}} I_n^\dagger + \frac{1}{\sqrt{(d_v^{(o)} + 1)}}U_{v, u_1}^\dagger \frac{1}{\sqrt{(d_v^{(o)} + 1)}}U_{v, u_1} + \frac{1}{\sqrt{(d_v^{(o)} + 1)}}U_{v, u_2}^\dagger \frac{1}{\sqrt{(d_v^{(o)} + 1)}} U_{v, u_2} \\
				& \hspace{4cm} + \dots + \frac{1}{\sqrt{(d_v^{(o)} + 1)}} U_{v, u_{d_v^{(o)}}}^\dagger \frac{1}{\sqrt{(d_v^{(o)} + 1)}} U_{v, u_{d_u^{(o)}}} = I_n.
			\end{split}
		\end{equation}
		It leads us to construct $(d_v^{(o)} + 1)$ coin operators for the vertex $v$ which are
		\begin{equation}\label{coin_operators}
			C_{v, u_i} = \frac{1}{\sqrt{(d_v^{(o)} + 1)}}U_{v, u_i} ~\text{for}~ i = 1, 2, \dots d_v^{(o)} ~\text{and}~ C_{v, v} = \frac{1}{\sqrt{(d_v^{(o)} + 1)}} I_n.
		\end{equation}
		The set of Kraus operators corresponding to the vertex $v$ is given by 
		\begin{equation}\label{Kraus_operators}
			\mathcal{C}_v = \left\{C_{v, u} = \frac{1}{\sqrt{(d_v^{(o)} + 1)}}U_{v, u}, u \in \Nbd(v) \right\} \cup \{C_{v, v}\},
		\end{equation}
		which we consider as the set of coin operators at vertex $v$, for the dynamics of quantum walk. 
		
		Recall the concept of standard basis states. Corresponding to  vertex $v$ we construct a quantum basis state vector $\ket{v} = (0, 0, \dots 0, 1(v\text{-th position}), 0, \dots 0)$. Also, corresponding to an edge $\overrightarrow{(v, u_i)}$ we construct a shift operator $S_{v, u_i} = \ket{u_i}\bra{v}$ for $i = 1, 2, \dots d_v^{(o)}$. As the walker may not move from vertex $v$ at time $t$, we consider $S_{v, v} = \ket{v}\bra{v}$ as a shift operator. The set of all shift operators at vertex $v$ is 
		\begin{equation}
			\mathcal{S}_v = \{S_{v, u}, u \in \Nbd(v)\} \cup \{S_{v, v}\}.
		\end{equation}
		
		The construction of $\mathcal{C}_v$ and $\mathcal{S}_v$ suggest that for every directed edge $\overrightarrow{(v, u)}$ there are a coin operator $C_{v, u}$ and a shift operator $S_{v, u}$. When we consider the evaluation on the graph $\overrightarrow{G}$, we need all of the operators corresponding to every directed edge. Combining the operators $C_{v, u}$ and $S_{v, u}$ for the edge $\overrightarrow{(v, u)}$ we get new set of Kraus operators $K_{v, u} = C_{v, u} \otimes S_{v, u}$. Note that 
		\begin{equation}
			\begin{split}
				\sum_{v \in V(G)} \sum_{u \in \Nbd(v)} K_{v, u}^\dagger K_{v, u} & = \sum_{v \in V(G)} \sum_{u \in \Nbd(v) \cup \{v\}} \left( C_{v, u}^\dagger \otimes S_{v, u}^\dagger \right) \left(C_{v, u} \otimes S_{v, u} \right)\\
				& = \sum_{v \in V(G)} \sum_{u \in \Nbd(v) \cup \{v\}} \left( C_{v, u}^\dagger \otimes \ket{v}\bra{u} \right) \left(C_{v, u} \otimes \ket{u}\bra{v} \right)\\
				& = \sum_{v \in V(G)} \left( \sum_{u \in \Nbd(v) \cup \{v\}} C_{v, u}^\dagger C_{v, u} \right) \otimes \ket{v}\braket{u | u}\bra{v}\\
				& = \sum_{v \in V(G)} I_n \otimes \ket{v}\bra{v} \hspace{.5cm} [\text{using equation (\ref{Kraus_condition_of_Weyl_operator_combination})}]\\
				&  = I_n \otimes I_n = I_{n^2}.
			\end{split} 
		\end{equation}
		Thus, 
		\begin{equation}
			\mathcal{K} = \{K_{v, u}: v \in V(G), u \in \Nbd(v) \cup \{v\}\}
		\end{equation}
		forms a set of Kraus operators on the graph $\overrightarrow{G}$. We use them to lead the movement of the quantum walker on $\overrightarrow{G}$. 
		
		Recall that, the objective of this work is building up a qPageRank algorithm. In the construction of classical PageRank we assumed that at $t = 0$ the internet surfer initiates surfing from any vertex chosen uniformly and randomly. We should consider this assumption, for developing our qPageRank algorithm, also. Hence, we define the initial density matrix as 
		\begin{equation}\label{initial_state}
			\rho^{(0)} = \sum_{v \in V(G)} \frac{1}{n^2}I_n \otimes \ket{v}\bra{v},
		\end{equation}
		where $I_n$ is identity matrix of order $n$. Note that $\rho^{(0)}$ is a density matrix of a quantum state because it is a positive semi-definite, Hermitian matrix and $\trace(\rho^{(0)}) = 1$. The probability of getting the walker at vertex $v$ is
		\begin{equation}
			p^{(0)}_v = \trace\left((I \otimes \ket{v}\bra{v}) \rho^{(0)} (I \otimes \ket{v}\bra{v}) \right) = \trace(\frac{1}{n^2}I_n \otimes \ket{v}\bra{v}) = \frac{1}{n^2}\trace(I_n) \times 1 = \frac{1}{n}.
		\end{equation}
		
		Now we discuss the change of the state of the walker at different time steps of discrete-time open quantum walk. Due to the particular tensor product structure of the initial state and Kraus operators, we can assume that the state of the walker can be expressed as 
		\begin{equation}\label{state_at_t}
			\rho^{(t)} = \sum_{v \in V(G)} \rho_v^{(t)} \otimes \ket{v}\bra{v},
		\end{equation}
		where $\rho_v^{(t)}$ is a positive semi-definite and Hermitian matrix, whose trace is not necessarily $1$. Recall that at $t = 0$ we have $\rho_v^{(0)} = \frac{I_n}{n^2}$, which is not a density matrix. But, $\rho^{(t)}$ is a density matrix for all $t$. The probability of getting the walker at the vertex $v$ at time $t$ is determined by
		\begin{equation}\label{probability_at_any_time_at_any_vertex}
			\begin{split}
				p^{(t)}_v & = \trace\left((I \otimes \ket{v}\bra{v}) \rho^{(t)} (I \otimes \ket{v}\bra{v}) \right) \\
				& = \trace\left((I \otimes \ket{v}\bra{v}) \left(\sum_{v \in V(G)} \rho_v^{(t)} \otimes \ket{v}\bra{v} \right) (I \otimes \ket{v}\bra{v}) \right)\\
				&  = \trace\left(\rho_v^{(t)} \otimes \ket{v}\bra{v} \right) = \trace\left(\rho_v^{(t)} \right) \times 1 = \trace\left(\rho_v^{(t)} \right).
			\end{split}
		\end{equation}
		
		The state of the walker at $(t + 1)$-th step governed by the Kraus operators in $\mathcal{K}$ is given by
		\begin{equation}\label{evolution_on_graph}
			\begin{split}
				\rho^{(t + 1)} & = \Psi_K(\rho^{(t)}) = \sum_{v \in V(G)} \sum_{u \in \Nbd(v) \cup \{v\}} K_{v, u} \rho_v^{(t)} \otimes \ket{v}\bra{v} K_{v, u}^\dagger \\
				& = \sum_{v \in V(G)} \sum_{u \in \Nbd(v) \cup \{v\}} C_{v, u} \rho_v^{(t)} C_{v, u}^\dagger \otimes S_{v, u} \ket{v}\bra{v} S_{v, u}^\dagger \\
				& = \sum_{v \in V(G)} \sum_{u \in \Nbd(v) \cup \{v\}} C_{v, u} \rho_v^{(t)} C_{v, u}^\dagger \otimes \ket{u}\braket{v | v}\braket{v | v}\bra{u} \\
				& = \sum_{v \in V(G)} \sum_{u \in \Nbd(v) \cup \{v\}} C_{v, u} \rho_v^{(t)} C_{v, u}^\dagger \otimes \ket{u}\bra{u}.
			\end{split}
		\end{equation}
		Recall that $u \in \Nbd(v)$ indicates that there is a directed edge $\overrightarrow{(v, u)}$ in the graph. Also, existence of the edge $\overrightarrow{(v, u)}$ indicates that $v \in B(u)$. Therefore, equation (\ref{evolution_on_graph}) can be written as 
		\begin{equation}\label{simplified_evolution}
			\begin{split}
				\rho^{(t + 1)} & = \sum_{v \in V(G)} \sum_{u \in \Nbd(v) \cup \{v\}} C_{v, u} \rho_v^{(t)} C_{v, u}^\dagger \otimes \ket{u}\bra{u} \\
				 & = \sum_{u \in V(G)} \sum_{v \in B(u) \cup \{u\}} C_{v, u} \rho_v^{(t)} C_{v, u}^\dagger \otimes \ket{u}\bra{u}\\
				 & = \sum_{v \in V(G)} \sum_{u \in B(v) \cup \{v\}} C_{u, v} \rho_u^{(t)} C_{u, v}^\dagger \otimes \ket{v}\bra{v}, ~\text{by interchanging the indices $u$ and $v$.}
			\end{split}
		\end{equation}
		Comparing equation (\ref{state_at_t}) and (\ref{simplified_evolution}) we observe that
		\begin{equation}\label{simplified_computation}
			\rho_v^{(t+1)} = \sum_{u \in B(v) \cup \{v\}} C_{u, v} \rho_u^{(t)} C_{u, v}^\dagger.
		\end{equation}
		
		The equation (\ref{simplified_computation}) reduces the size of Hilbert space in our consideration. In general, the dynamics of our quantum walk formally takes place in a Hilbert space of dimension $n^2$. But, because of the separable Kronecker product structure of the states and operators involved, one can restrict the actual computations to a Hilbert subspace of dimension $n$. This is a known behavior of discrete-time quantum walks on graphs.\cite{paparo2012google, paparo2013quantum, paparo2014quantum}. The state $\ket{v}\bra{v}$ in equation (\ref{state_at_t}) remains unchanged in all the steps. We update the coin space corresponding to a vertex from time to time using equation (\ref{simplified_computation}). Therefore, the Hilbert space for our calculations is  $\mathcal{H}^{n}$, which is a space of dimension $n$.

	\section{Kraus operators to restart the walking in discrete-time open quantum walk}
	\label{section_4}
		
		In subsection \ref{PageRank}, we discussed that the movement of a surfer may not be limited to the hyperlinked vertices. The random walker can move to any vertex of the graph in the next time instance and restart surfing from there. To model this phenomenon using open quantum walk we also develop a set of Kraus operators. Given any vertex $u$ we define $n$ operators $S_{uv} = \ket{u}\bra{v}$, such that 
		\begin{equation}
			S_{uv} \ket{v}\bra{v} S_{uv}^\dagger = \ket{u}\braket{v | v} \braket{v | v} \bra{u} = \ket{u}\bra{u}.
		\end{equation}
		Also, we have
		\begin{equation}
			\sum_{v \in V(G)} S_{uv}^\dagger S_{uv} = \sum_{v \in V(G)} \ket{v}\bra{v} = I_n.
		\end{equation}
		Considering all possible pairs of the vertices $u, v$ in the graph we have $n^2$ operators $S_{u, v}$. We define a set of operators 
		\begin{equation}\label{damping_operators}
			\mathcal{D} = \left\{D_{u, v}: D_{u, v} = \frac{1}{\sqrt{n}}(I_n \otimes S_{u, v}) \right\}.
		\end{equation}
		The set $\mathcal{D}$ is a set of Kraus operators because
		\begin{equation}
			\begin{split}
				\sum_{v \in V(G)} D_{u, v}^\dagger D_{u, v} & =  \sum_{v \in V(G)} \frac{1}{\sqrt{n}}(I_n \otimes S_{u, v})^\dagger \frac{1}{\sqrt{n}}(I_n \otimes S_{u, v}) = \frac{1}{n} \sum_{v \in V(G)} (I_n \otimes S_{u, v}^\dagger S_{u, v}) \\
				& = \frac{1}{n} \left(I_n \otimes \sum_{v \in V(G)} S_{u, v}^\dagger S_{u, v} \right) = \frac{1}{n} I_n \otimes I_n = \frac{I_{n^2}}{n} \\
				\text{or}~ \sum_{u \in V(G)} \sum_{v \in V(G)} D_{u, v}^\dagger D_{u, v} & =  \sum_{u \in V(G)} \frac{I_{n^2}}{n} = I_{n^2}.
			\end{split}
		\end{equation}
	
		Now applying the operators of $\mathcal{D}$ on $\rho^{(t)}$ mentioned in equation (\ref{state_at_t}) we get
		\begin{equation}\label{damping_channel_application}
			\begin{split}
				\rho^{(t + 1)}_D & = \Psi_D(\rho^{(t)}) = \sum_{u \in V(G)} \sum_{v \in V(G)} D_{u, v} \rho^{(t)} D_{u, v}^\dagger \\
				& = \sum_{u \in V(G)} \sum_{v \in V(G)} \frac{1}{\sqrt{n}}(I_n \otimes S_{u, v}) \left(\sum_{v \in V(G)} \rho_v^{(t)} \otimes \ket{v}\bra{v} \right) \frac{1}{\sqrt{n}}(I_n \otimes S_{u, v})^\dagger \\
				& = \sum_{u \in V(G)} \frac{1}{n} \sum_{v \in V(G)} \rho_v^{(t)} \otimes S_{u, v} \ket{v}\bra{v} S_{u, v}^\dagger \\
				& = \sum_{u \in V(G)} \left(\frac{1}{n} \sum_{v \in V(G)} \rho_v^{(t)}\right) \otimes \ket{u}\bra{u}.
			\end{split}
		\end{equation}
		Note that we find an average of all matrices $\rho_v^{(t)}$ at the vertex $u$ after application of the operators in $\mathcal{D}$. Also, the dimension of the evaluation is again limited to the coin space $\mathcal{H}^n$. Equation (\ref{damping_channel_application}) assists us to restart the walker during internet surfing as discussed in subsection \ref{PageRank}.

	\section{qPageRank of the vertices of a graph}
	
		Till now we have demonstrated two types of evaluations on the state $\rho^{(t)}$ in Section \ref{section_3} and Section \ref{section_4}. We have mentioned two types of Kraus operators in equation (\ref{evolution_on_graph}) of Section \ref{section_3} and equation (\ref{damping_channel_application}) of Section \ref{section_4} which we utilize as the coin operators of the discrete-time quantum walk. The operators in equation (\ref{evolution_on_graph}) assists the walker to move along a directed edge. Also, the operators in equation (\ref{damping_channel_application}) helps the walker to be teleported  to any vertex in the graph for initiating the quantum walk again. In the construction of Google's PageRank we combine these two processes, which we have discussed in equation (\ref{Google_matrix}). To develop qPagerank also, we combine these processes to govern the dynamics of the walker. Recall that $\alpha$ is the PageRank parameter, such that $(1 - \alpha)$ is the probability of teleportation. Then, 
		\begin{equation}\label{evolution_with_damping_factor}
			\begin{split}
				\rho^{(t + 1)} & = \alpha \sum_{u \in V(G)} \sum_{v \in B(u) \cup \{u\}} C_{v, u} \rho_v^{(t)} C_{v, u}^\dagger \otimes \ket{u}\bra{u} + (1 - \alpha) \sum_{u \in V(G)} \left(\frac{1}{n} \sum_{v \in V(G)} \rho_v^{(t)}\right) \otimes \ket{u}\bra{u} \\
				& = \sum_{u \in V(G)} \left(\alpha \sum_{v \in B(u) \cup \{u\}} C_{v, u} \rho_v^{(t)} C_{v, u}^\dagger\right) \otimes \ket{u}\bra{u} + \sum_{u \in V(G)} \left(\frac{(1 - \alpha)}{n} \sum_{v \in V(G)} \rho_v^{(t)}\right) \otimes \ket{u}\bra{u} \\
				& = \sum_{u \in V(G)} \left[\alpha \sum_{v \in B(u) \cup \{u\}} C_{v, u} \rho_v^{(t)} C_{v, u}^\dagger + \frac{(1 - \alpha)}{n} \sum_{v \in V(G)} \rho_v^{(t)}\right] \otimes \ket{u}\bra{u}\\
				& = \sum_{u \in V(G)} \rho_u^{(t + 1)} \otimes \ket{u}\bra{u},
			\end{split}
		\end{equation}
		where
		\begin{equation}\label{classical_evolution_with_damping_factor}
			\rho_u^{(t + 1)} = \alpha \sum_{v \in B(u) \cup \{u\}} C_{v, u} \rho_v^{(t)} C_{v, u}^\dagger + \frac{(1 - \alpha)}{n} \sum_{v \in V(G)} \rho_v^{(t)}.
		\end{equation}
		The initial state $\rho^{(0)}$ is mentioned in equation ($\ref{initial_state}$).
		
		In equation (\ref{evolution_with_damping_factor}), the state $\ket{u}\bra{u}$ remains unchanged in every time instance. We only update the coin space corresponding to a vertex from time to time using equation (\ref{classical_evolution_with_damping_factor}). Therefore, an $n$ dimensional Hilbert space is sufficient for our calculation. A classical computer can perform the matrix multiplications and additions which are required in equation (\ref{classical_evolution_with_damping_factor}) in polynomial time.
		
		Recall from equation (\ref{probability_at_any_time_at_any_vertex}) that the probability of getting the walker at vertex $u$ at time $(t + 1)$ is given by
		\begin{equation}\label{qPageRank}
			p_u^{(t+1)} = \trace \rho_u^{(t + 1)}.
		\end{equation}
		
		Equation (\ref{evolution_with_damping_factor}) provides the time evolution of the surfer on the internet graph, who follows discrete-time open quantum walk. In every time instance, the probability of getting the walker at different vertices becomes updated and converges to a limit as the dynamics is Markovian. We discuss this convergence in further details with numerical experiments with the samples of vertices belonging to different graphs, in Subsection \ref{Convergence_test_of_qPageRank}. We consider the limiting probability distribution as our qPageRank of the vertices. Hence, we have the definition below.
		\begin{definition}\label{qPageRank_definiiton}
			We define the qPageRank vector at time $t$ for a graph $G$ as $q\Pi^{(t)} = (p_1^{(t)}, p_2^{(t)}, \dots, p_n^{(t)})$. The qPageRank vector of a graph $G$ is
			\begin{equation}
				q\Pi = \lim_{t \rightarrow \infty} q\Pi^{(t)} = \lim_{t \rightarrow \infty} (p_1^{(t)}, p_2^{(t)}, \dots, p_n^{(t)}),
			\end{equation}
			where $p_u^{(t)}$ is the probability of getting the walker at vertex $u$ at time $t$ determined by equation (\ref{qPageRank}).
		\end{definition}
		In practice, instead of calculating $t \rightarrow \infty$, we compute $q\Pi^{(t)}$ for a finite number of iterations on $t$ to achieve the limiting probability distribution. To determine the number of required iterations we fix a stopping criterion. Here, we need the distance between two qPageRank vectors. The Euclidean distance between $q\Pi^{(T - 1)}$ and $q\Pi^{(T)}$ in two consecutive steps $t = (T - 1)$ and $t = T$ is determined by \cite{horn2012matrix}
		\begin{equation}
			||q\Pi^{(T)} - q\Pi^{(T - 1)}|| = \sqrt{\left(p_1^{(T)} - p_1^{(T - 1)} \right)^2 + \left(p_2^{(T)} - p_2^{(T - 1)} \right)^2 + \dots + \left(p_n^{(T)} - p_n^{(T - 1)} \right)^2}.
		\end{equation}
		We stop computing $q\Pi^{(t)}$ at $t = T$, if the distance
		\begin{equation}\label{stopping_creterion}
			||q\Pi^{(T)} - q\Pi^{(T - 1)}|| < \epsilon.
		\end{equation}
		Here $\epsilon$ is a small number, which we consider as a numerical error. Numerically, we can accept $q\Pi^{(T)}$ as the value of $q\Pi$. In this article, we consider $\epsilon = .0001$, which generates the correct value of $q\Pi$ up to $3$ decimal places.
		 
		Below, we mention a summary of steps involved in our algorithm for calculating qPageRank of the vertices in any directed graph. We also mention the steps of the calculations using a classical computer.
		\begin{procedure}\label{qPageRank_algo}
			Follow the below steps to calculate the qPageRank of the vertices in a given graph:
			\begin{enumerate}
				\item 
					\textbf{Preparation of initial state:} The initial state $\rho^{(0)}$ of the system to be prepared following equation (\ref{initial_state}). Classically, we make a list $\mathcal{L}$ of length $n$ with the matrices $\frac{I_n}{n^2}$ corresponding to the vertices. 
				\item 
					\textbf{Construction of operators:} Corresponding to every directed edge in the graph, we construct the coin operators following equation (\ref{coin_operators}).
				\item 
					\textbf{Quantum evolution:} We evaluate the quantum state iteratively following equation (\ref{evolution_with_damping_factor}). Classically, to reduce the size of our calculations, we apply equation (\ref{classical_evolution_with_damping_factor}) to update elements in the list $\mathcal{L}$. 
				\item 
					\textbf{qPageRank measurement:} We calculate the qPageRank of a vertex by measuring the state following the equation (\ref{qPageRank}). Classically, we calculate the trace of the matrices in $\mathcal{L}$ at different time steps to get $q\Pi^{(t)}$ mentioned in Definition \ref{qPageRank_definiiton}. We fix an acceptable numerical error $\epsilon$. We terminate the process when the stopping criterion mentioned in equation (\ref{stopping_creterion}) is satisfied. The qPageRank of a vertex is the trace of the corresponding matrix in the list $\mathcal{L}$.
			\end{enumerate}	
		\end{procedure}
		
		If the graph is undirected, we can assign two opposite orientations on every edge and make it a directed graph. Hence, we can apply the above procedure for calculating the qPageRank of the vertices in an undirected graph.
		
		Here, we discuss a significant benefit of utilizing the discrete-time open quantum walk for determining qPageRank. We have already mentioned that there are multiple proposals of qPageRank depending on various quantum walks; for instance, the qPageRank based on discrete-time quantum walks \cite{paparo2012google}, qPageRank based on continuous-time open quantum walk \cite{sanchez2012quantum, boito2023ranking}, qPageRank based on Szegedy's quantum walk \cite{paparo2013quantum, paparo2014quantum, chawla2020discrete}, and \cite{ortega2023generalized}, etc. The quantum walk dynamics of the discrete-time quantum walk, the continuous-time open quantum walk, and Szegedy's quantum walk are led by unitary matrices. At every step of the quantum walk, they calculate an instantaneous qPageRank $q\Pi^{(t)}$. The unitary matrix governing the quantum walk prevents the convergence of the instantaneous qPageRank sequence. To assure convergence, we consider the time average of instantaneous qPageRank, which is $\{\frac{1}{T} \sum_{t = 0}^T q\Pi^{(t)}\}_{T = 0}^\infty$ \cite{aharonov2001quantum}. In these methods, $\lim_{T \rightarrow \infty} \frac{1}{T} \sum_{t = 0}^T q\Pi^{(t)}$ is considered as qPageRank. In our case, we involve the Markovian dynamics, which assures the convergence of $q\Pi^{(t)}$ to a limit for $t \rightarrow \infty$. Therefore, we do not need the sequence of time averages. It reduces the calculations and makes the process faster.

	\section{qPageRanks of nodes on various networks}
	
		In this section, we present our observations from the numerical experiments with Procedure \ref{qPageRank_algo} on a number of graphs and networks. We also compare the qPageRank with the PageRank for both directed and undirected graphs. Our construction of qPageRank is applicable to directed graphs. Recall that we can represent an undirected graph as a directed graph by considering two opposite directions on every edge. We plot bar diagrams to represent the importance of different vertices based on PageRank and qPageRank. In all the bar diagrams, the blue bars represent classical PageRank and the orange bars represent qPageRank. Our computations utilize the Python library \textit{NetworkX} \cite{hagberg2008exploring}.
		
		The PageRank and qPageRank may behave differently depending on the graphs. To compare them we try to realize how the ranks of the vertices generated by these two methods are correlated. We rank the vertices of the graph with the integers $\{1, 2, \dots n\}$ based on PageRank and qPageRank values separately. We break the ties arbitrarily. Then we calculate Kendall's rank correlation coefficient \cite{kendall1938new}, if the ranking on the vertices mismatch.  
		
		\subsection{Comparison between PageRank and qPageRank for a few simple graphs}
			
\begin{figure}[h!]
				\centering
				\begin{subfigure}[b]{0.4\textwidth}
					\includegraphics[scale=.5]{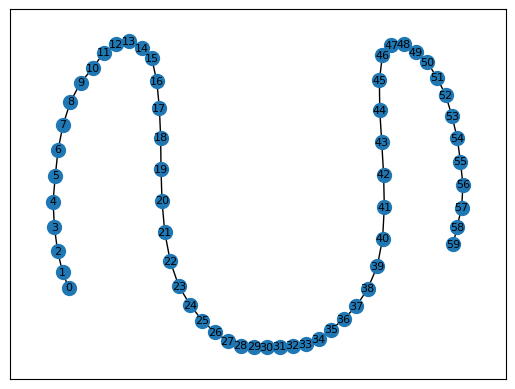}
					\caption{The Path graph with $60$ vertices.}
					\label{path_10}
				\end{subfigure}
				\begin{subfigure}[b]{0.56\textwidth}
					\includegraphics[width=\textwidth]{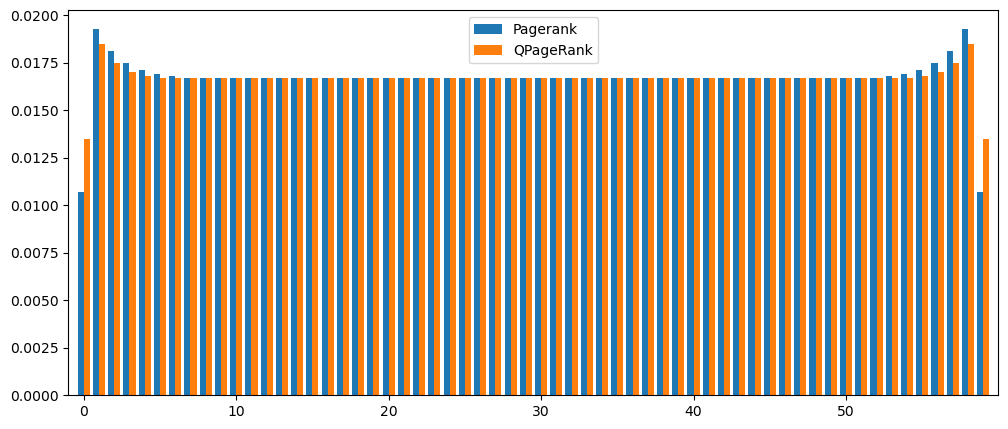}
					\caption{The blue and orange bars represent the qPageRank and PageRank of different vertices of a path graph.}
					\label{path_bar}
				\end{subfigure}
				\caption{Comparison between the PageRank and qPageRank of vertices on path graph with $60$ vertices. (Color online)}
				\label{path}
			\end{figure}

			For simplicity, we initiated our study with a few simple graphs. The first one is a path graph with sixty vertices which is depicted in figure \ref{path_10}. The PageRank and qPageRank of different vertices are mentioned in the table \ref{path_graph_values}.
			\begin{table}[h!]
				\centering
				\begin{tabular}{| c | c | c | c | c | c | c | c | c | c |}
					\hline
					Vertices & 0 \& 59 &	1 \& 58 &	2 \& 57 &	3 \& 56 &	4 \& 55 &	5 \& 54 &	6 \& 53 &	7 \& 52 &	others\\
					\hline 
					PageRank &	0.0107 &	0.0193 &	0.0181 &	0.0175 &	0.0171 &	0.0169 &	0.0168 &	0.0167 &	0.0167\\
					\hline 
					qPageRank &	0.0135 &	0.0185 &	0.0175 &	0.0170 &	0.0168 &	0.0167 &	0.0167 &	0.0167 &	0.0167\\
					\hline 
				\end{tabular}
				\caption{Values of PageRank and qPageRank of a path graph with $60$ vertices.}
				\label{path_graph_values}
			\end{table}
			
			We represent this data as a bar diagram in figure \ref{path_bar}. We place the vertices in $X$ axis as well as the PageRanks and qPageRanks in $Y$ axis.
		
			There are a few observations from the data about the ranking of vertices on the path graphs. The values of PageRank and qPageRank are different only for the first and last six vertices at the two ends. For the other vertices, the values of PageRank and qPageRank are equal to $0.0167$. Although the values of PageRank and qPageRank are different for a few vertices, the rankings of the vertices in both methods are identical. The extreme vertices $0$ and $59$ have an equal ranking based on PageRank and qPageRank. Similarly, the pairs of vertices $(1, 58), (2, 57), (3, 56)$ and $(4, 55)$ have equal ranking using both methods. The rankings based on PageRank and qPageRank are also identical. Also, note that the difference between the values of PageRank of vertices $0$ and $1$. The difference between qPageRank values of these vertices is less, comparatively. 
			
			In addition, we compute PageRank and qPageRank for a number of undirected graphs, for instance complete graphs, cycle graphs, star graphs, wheel graphs, and balanced trees for our investigations. Table \ref{undirected_graph_results} contains the bar diagrams representing the PageRank and qPageRank of different graphs. We can observe a few interesting characteristics of qPageRank from our computation, which we mention below.
			
			\begin{table} 
				\begin{tabular}{|p{1.5cm}| c | c |}
					\hline 
						\textbf{Graphs} & \textbf{Figure} & \textbf{Comparison between PageRank and qPageRank} \\
					\hline
						Complete graph with 20 vertices & \begin{minipage}{.3\textwidth}  \includegraphics[scale = .4]{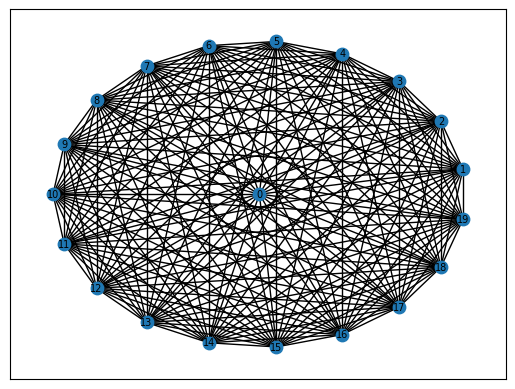} \end{minipage} & \begin{minipage}{.5\textwidth}  \includegraphics[height = 3.5cm, width = 8.5cm]{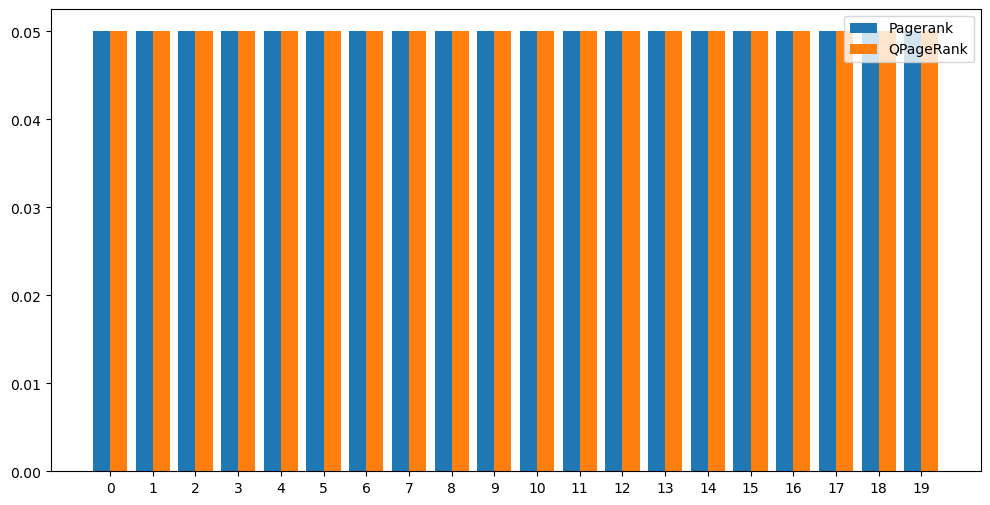} \end{minipage}\\
					\hline 
						Cycle Graph with 60 vertices & \begin{minipage}{.3\textwidth}  \includegraphics[scale = .4]{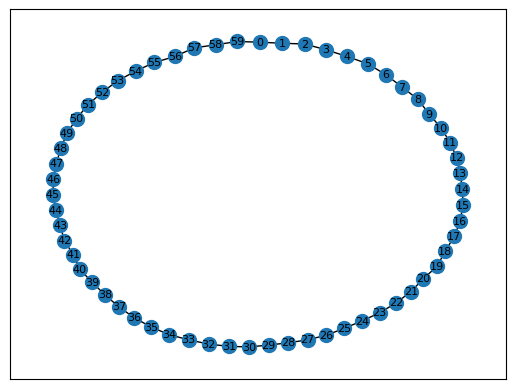} \end{minipage} & \begin{minipage}{.5\textwidth}  \includegraphics[scale = .35]{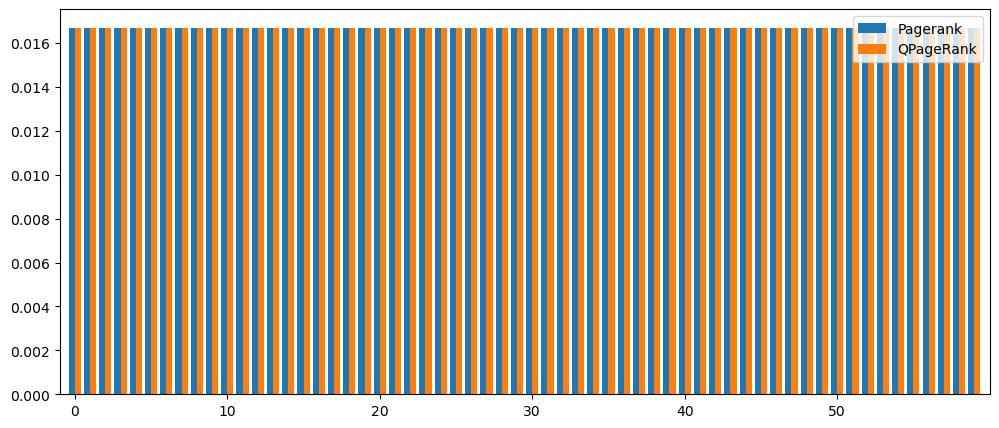} \end{minipage}\\
					\hline
						Star Graph with 61 vertices & \begin{minipage}{.3\textwidth}  \includegraphics[scale = .4]{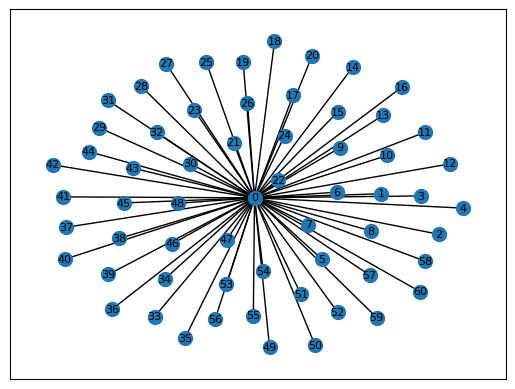} \end{minipage} & \begin{minipage}{.5\textwidth}  \includegraphics[scale = .35]{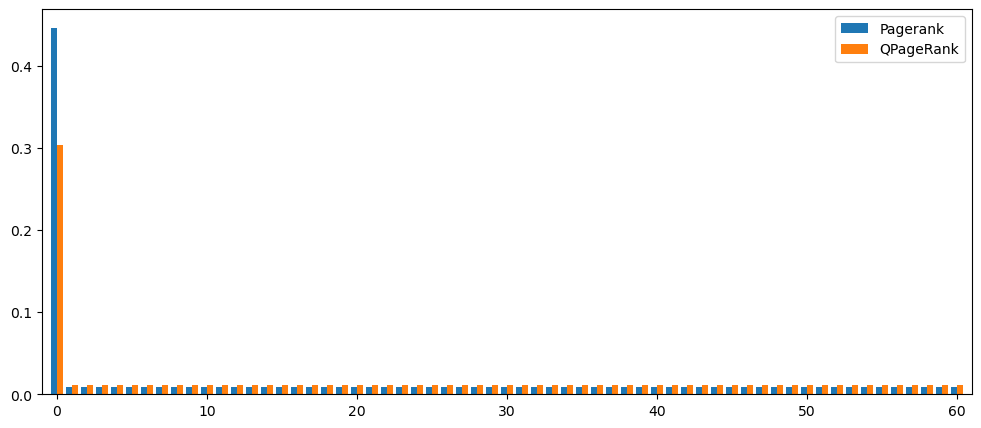} \end{minipage}\\ 
					\hline
						Wheel graph with 61 vertices & \begin{minipage}{.3\textwidth}  \includegraphics[scale = .4]{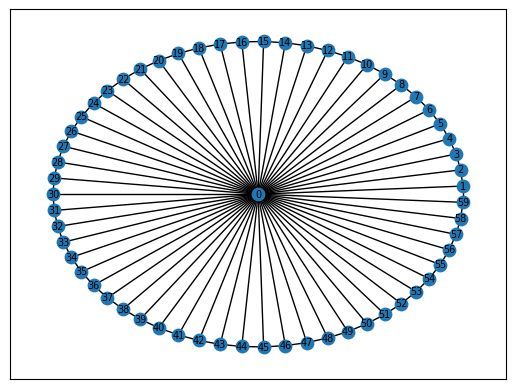} \end{minipage} & \begin{minipage}{.5\textwidth}  \includegraphics[scale = .35]{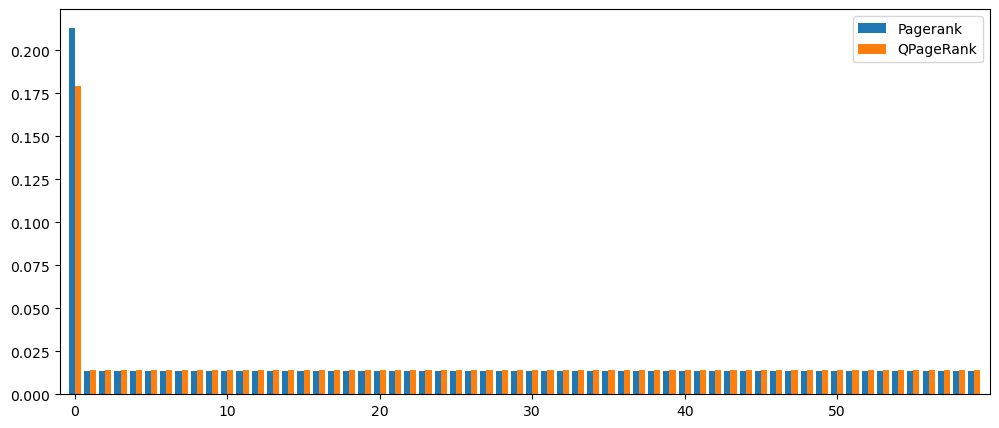} \end{minipage}\\
					\hline
				\end{tabular}
				\caption{Comparison between PageRank and qPageRank of the vertices in undirected graphs with symmetry. (Color online)}
				\label{undirected_graph_results}
			\end{table} 
			
			\begin{figure}[h!]
				\begin{subfigure}[b]{.49\textwidth}
					\centering
					\includegraphics[scale = .55]{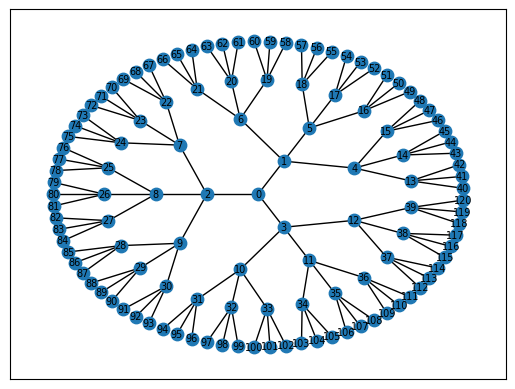}
					\caption{Undirected balanced tree $(3, 4)$ with $121$ vertices which are distributed into four layers surrounding the central vertex $0$.}
					\label{balanced_tree_3_4} 
				\end{subfigure}
				\hspace{.5cm}
				\begin{subfigure}[b]{.49\textwidth}
					\centering
					\includegraphics[height = 5cm, width = 8cm]{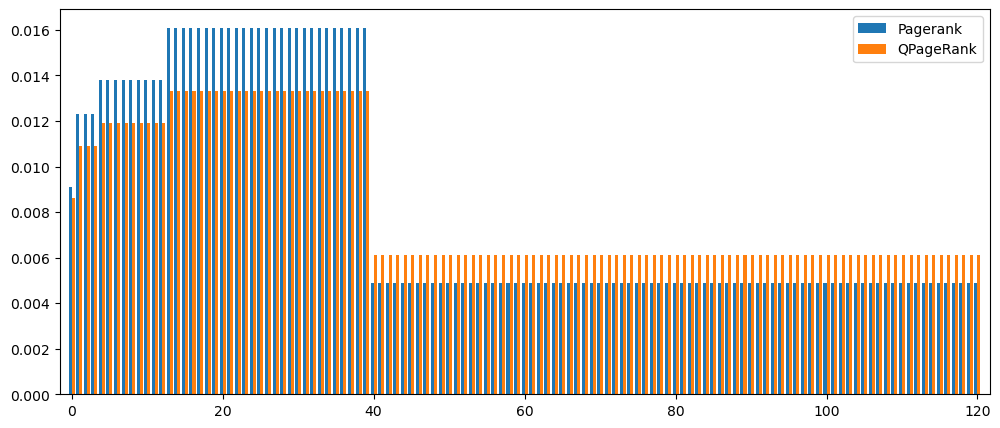}
					\caption{PageRank and qPageRank of the vertices in undirected balanced tree $(3, 4)$ is represented with bars.}
					\label{balanced_tree_3_4_bar}
				\end{subfigure}
				\caption{Undirected balanced tree as well as a bar diagram representing the PageRank and qPageRank of its vertices. (Color online)}
			\end{figure}
			
			In Table \ref{undirected_graph_results}, we consider a complete graph with 20 vertices, a cycle graph with $60$ vertices, as well as a stat graph and a wheel graph each of 61 vertices. We observe that for the cycle and complete graphs, the values of qPageRank exactly match with PageRank for every vertex. These values are $.05$ and $0.0167$, respectively, for the complete graph, and the cycle graph, under our consideration. Also, increasing the number of vertices we observe the same phenomenon. For the other graphs discussed in this article the values of PageRank and qPageRank do not match.
			
			For the star graphs, wheel graphs, balanced tree, etc. the qPageRank reflects the symmetry in graphs. For our numerical investigations, we consider a star graph with $61$ vertices. The vertex $0$ acts as a central vertex. There are $60$ vertices in the surroundings. The central vertex has maximum values of PageRank and qPageRank. The PageRank and qPageRank of the other vertices in the surrounding are equal, which are mentioned in Table \ref{Path_and_Wheel_values}. The difference in the PageRank values between the central and surrounding vertices is greater than the corresponding difference in qPageRank values.
			\begin{table}[h!]
				\centering
				\begin{tabular}{|c | c | c | c |}
					\hline 
					Graphs & Vertices & PageRank & qPageRank \\
					\hline 
					Path Graph & Central vertex ($0$) & $0.4462$ & $0.3029$\\
					\cline{2-4}
					& Surrounding vertices ($1$ to $60$) & $0.0092$ & $0.0116$\\
					\hline 
					Wheel graph & Central vertex ($0$) & $0.2132$ & $0.1794$\\
					\cline{2-4}
					& Surrounding vertices ($1$ to $60$) & $0.0133$ & $0.0139 $\\
					\hline 
				\end{tabular}
				\caption{Values of PageRank and qpageRank of Path and Wheel graphs.}
				\label{Path_and_Wheel_values}
			\end{table}
			 
			Consider the balanced tree $(3, 4)$ with $121$ vertices which is depicted in Figure \ref{balanced_tree_3_4}. The vertices of a balanced tree are distributed into four layers surrounding the central vertex $0$. Vertices in the 1st layer surrounding vertex $0$ are $1, 2$, and $3$. In layer $2$ the vertices are from $4$ to $12$. In layer $3$ the vertices are from $13$ to $39$. The other vertices from $40$ to $120$ live in the outermost layer. Every vertex other than the vertices in the outermost layer has three neighbors in the immediately next layers. We observe that the values of PageRank and qPageRank are equal for all the vertices in same layer. These values are presented in Table \ref{Balanced_tree_values}. 
			\begin{table}[h!]
				\centering 
				\begin{tabular}{| c | c | c |}
					\hline 
					Vertices in layer & PageRank & qPageRank \\
					\hline
					Central vertex $0$ & $0.0091$ & $0.0086$ \\ 
					\hline
					Vertices in the first layer ($0, 1, 2$) & $0.0123 	$ & $0.0109$\\
					\hline	
					Vertices in the second layer ($4, 5, \dots 12$)& $0.0138$ & $0.0119$ \\
					\hline
					Vertices in the third layer ($13, 14, \dots 39$) & $0.0161$ & $0.0133$ \\
					\hline 
					Vertices in the fourth layer ($40, 41, \dots 120)$ & $0.0049$ & $0.0061$\\
					\hline   
				\end{tabular}
				\caption{Values of PageRank and qPageRank of undirected Balanced Tree $(3, 4)$}
				\label{Balanced_tree_values}
			\end{table}
			
			We also consider a number of undirected graphs without any trivial structure of symmetry, which are the Karate Club \cite{zachary1977information} network, Watts-Strogatz graph \cite{watts1998collective}, as well as Barabási and Albert graph \cite{barabasi1999emergence}. The last two are random graphs. Existence of an edge in these graphs depend on probability of a random variable. Therefore, they are not unique. Two specific instances of these graphs are considered for our investigations. The comparison between PageRank and qPageRank of different vertices in these graphs are presented by bar diagrams in Table \ref{special_undirected_graph_results}.
			\begin{table}
				\centering
				\begin{tabular}{|p{1.5cm}| c | p{7.5cm} |p{1.5cm}|}
					\hline 
						\textbf{Graphs} & \textbf{Figure} & \textbf{Comparison between PageRank and qPageRank} & \textbf{Rank correlation}\\
					\hline
						Karate Club with $34$ vertices and $78$ edges & \begin{minipage}{.3\textwidth}  \includegraphics[scale = .4]{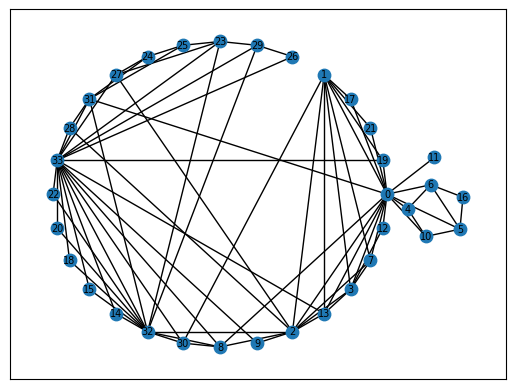} \end{minipage} & \begin{minipage}{.5\textwidth}  \includegraphics[height = 3.5cm, width = 7.5cm]{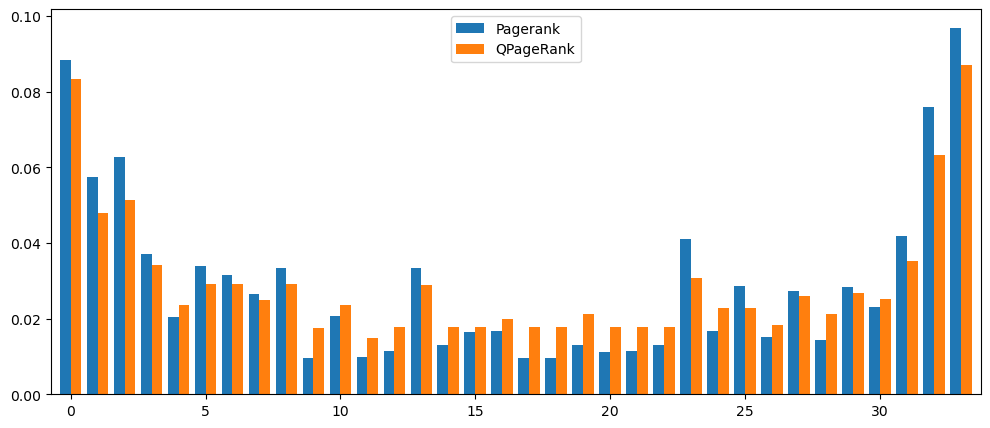} \end{minipage} & $0.860963$\\
					\hline
						Watts-Strogatz network with $80$ vertices and $80$ edges & \begin{minipage}{.3\textwidth}  \includegraphics[scale = .4]{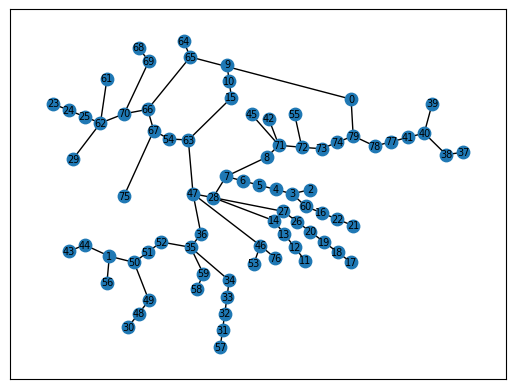} \end{minipage} & \begin{minipage}{.5\textwidth}  \includegraphics[height = 3.5cm, width = 7.5cm]{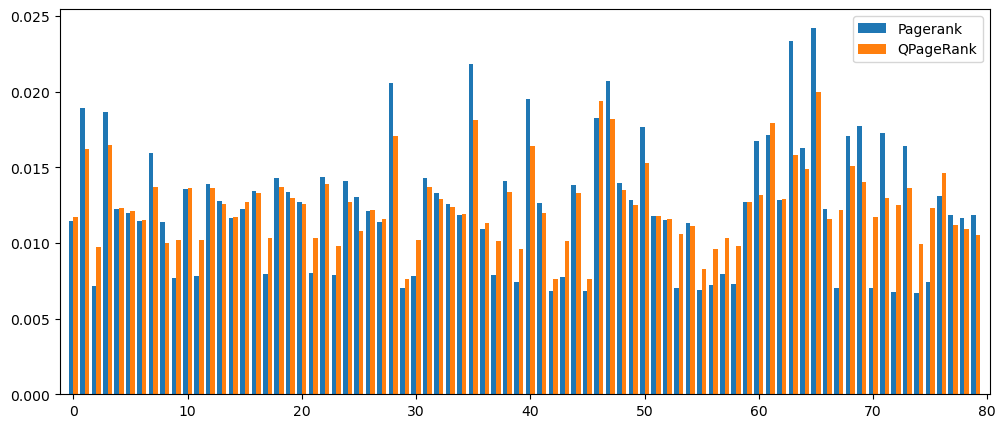} \end{minipage} & $0.6578$\\
					\hline
						Barabási and Albert network with 120 vertices and $464$ edges & \begin{minipage}{.3\textwidth}  \includegraphics[scale = .4]{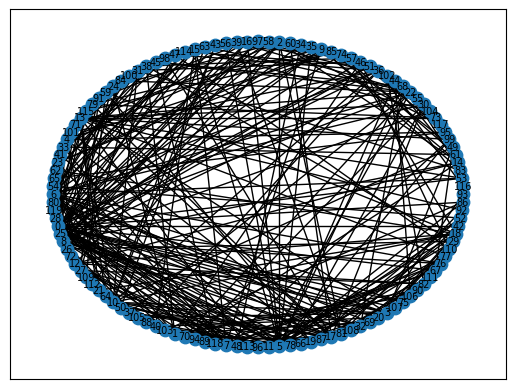} \end{minipage} & \begin{minipage}{.5\textwidth}  \includegraphics[height = 3.5cm, width = 7.5cm]{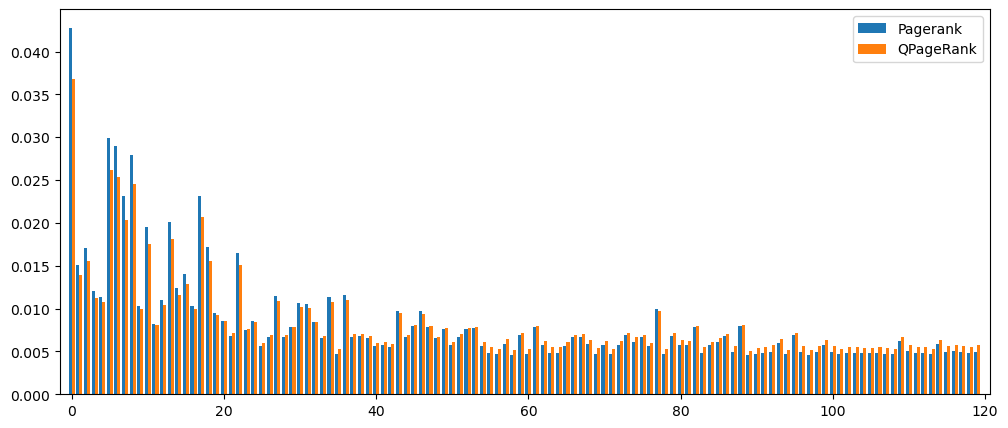} \end{minipage} & $0.9986$\\
					\hline
				\end{tabular} 
				\caption{Comparison between PageRank and qPageRank of the vertices in undirected graphs with symmetry. The PageRank and qPageRank may behave differently for these graphs. (Color online.)}
				\label{special_undirected_graph_results}
			\end{table}	
			The ranking of vertices generated by PageRank and qPageRank do not exactly match for these graphs. Therefore, we calculate the rank correlation between them, which are also collected in Table \ref{special_undirected_graph_results}.

		\subsection{Comparison between PageRank and qPageRank for a few directed graphs}
			
			Now, we consider the directed graphs, where most edges have only one direction. There are a few graphs for which PageRank and qPageRank act similarly, such as the directed trinary tree, GN, GNC, and GNR networks \cite{krapivsky2005network}. For many others PageRank and qPageRank behave differently, for instance, the scale-free network, Erd\H{o}s-R\'enyi random graph, random\_k\_out graph, etc. We initiate our discussion with the graphs for which PageRank and qPageRank match.
			
\begin{figure}
				\centering 
				\begin{subfigure}[a]{.5\textwidth}
					\centering
					\includegraphics[scale = .6]{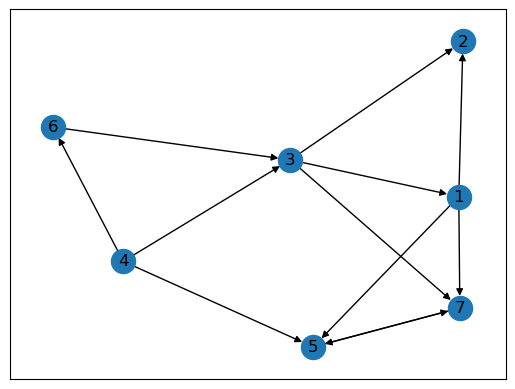}
					\caption{A directed graph which does not belong to any specific class of graphs, considered in \cite{paparo2013quantum}.}
					\label{Paparo_graph}
				\end{subfigure}
				\hspace{.25cm} 
				\begin{subfigure}[a]{.44\textwidth}
					\centering
					\includegraphics[width=8cm, height = 6cm]{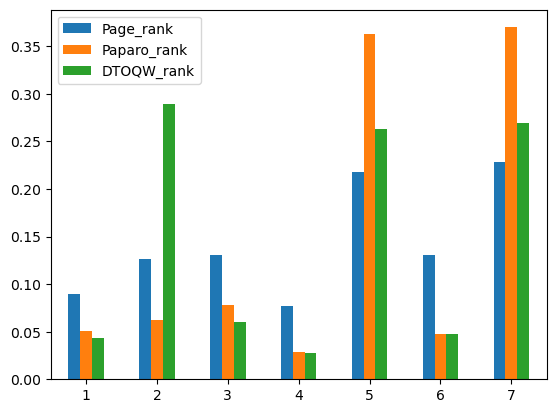}
					\caption{PageRank and qPageRank of the vertices in directed trinary tree platted as a bar diagram.}
					\label{Paparo_graph_bar}
				\end{subfigure}
				\caption{We consider a directed graph with $7$ vertices and $11$ edges which are depicted in subfigure \ref{Paparo_graph}. For the vertices we plot the PageRank, qPageRank based on \cite{paparo2013quantum} and qPageRank based on DTOQW with bar diagrams in subfigure \ref{Paparo_graph_bar}. (Color online.)}
				\label{Paparo} 
			\end{figure}

			We first compute our qPageRank for a directed graph depicted in Figure \ref{Paparo_graph}, which was also considered in \cite{paparo2013quantum} earlier. This graph does not belong to any particular class of graphs. We find that the ranking generated by DTOQW is different from the PageRank and the qPageRank developed in \cite{paparo2013quantum}. The values of DTOQW-based qPageRank for different vertices are available in Table \ref{Paparo_graph_table}. Also, Table \ref{Paparo_graph_table} provides a comparison between different rankings. Different PageRank values for the vertices are represented by a multiple-bar diagram in Figure \ref{Paparo_graph_bar}.
			\begin{table}[h!]
				\centering 
				\begin{tabular}{|p{1.5cm}| p{2cm} | p{2cm} | p{2cm} | p{2cm} |}
					\hline 
					\textbf{Vertices as labeled in Figure \ref{Paparo_graph}}  & qPageRank based on DTOQW & Ranking based on DTOQW & Ranking based on qPageRank developed in \cite{paparo2013quantum}  & Ranking based on PageRank \cite{paparo2013quantum} \\
					\hline
					1 & 0.0434 & 6 & 5 & 6 \\
					\hline
					2 & 0.2895 & 1 & 4 & 5 \\
					\hline
					3 & 0.0601 & 4 & 3 & 4 \\
					\hline
					4 & 0.0272 & 7 & 7 & 7 \\
					\hline
					5 & 0.2627 & 3 & 2 & 2 \\
					\hline
					6 & 0.0473 & 5 & 6 & 3 \\
					\hline
					7 & 0.2697 & 2 & 1 & 1 \\
					\hline
				\end{tabular}
				\caption{PageRank and qPageRank values for a toy graph depicted in Figure \ref{Paparo_graph}.}
				\label{Paparo_graph_table}
			\end{table}
			
\begin{figure}
				\centering 
				\begin{subfigure}[a]{.5\textwidth}
					\centering
					\includegraphics[scale = .6]{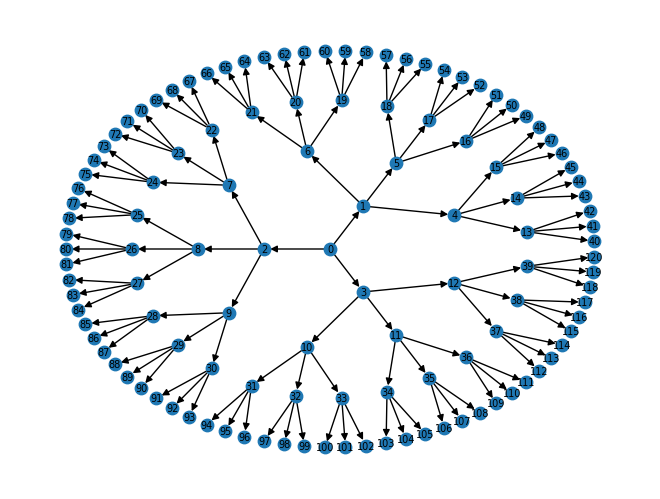}
					\caption{A directed trinary tree, where the edges are directed outwards. There is a central vertex marked as $0$ and four layers of vertices surrounding it}
					\label{trinary_tree}
				\end{subfigure}
				\hspace{.25cm} 
				\begin{subfigure}[a]{.44\textwidth}
					\centering
					\includegraphics[width=4cm, height = 6cm]{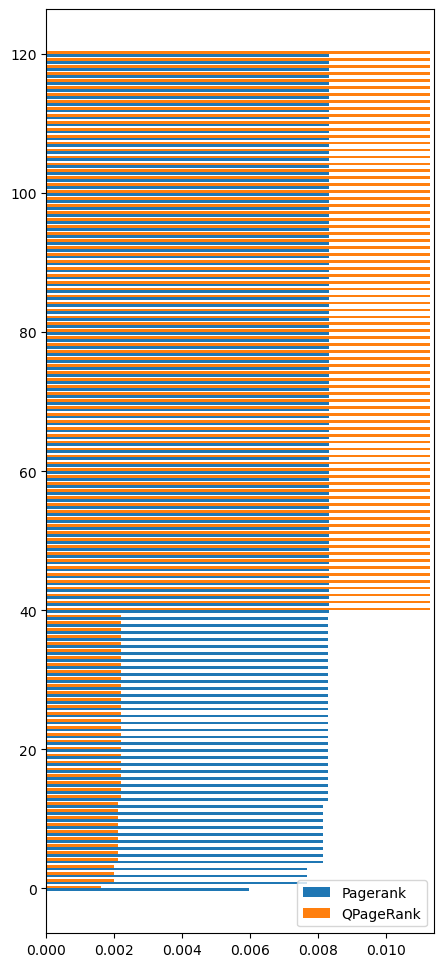}
					\caption{PageRank and qPageRank of the vertices in directed trinary tree platted as a bar diagram.}
					\label{trinary_tree_bar}
				\end{subfigure}
				\caption{We consider a directed trinary tree with $121$ vertices and $120$ edges which are depicted in subfigure \ref{trinary_tree}. For the vertices we plot the PageRank and qPageRank as bar diagram in subfigure \ref{trinary_tree_bar}. Note that, the qPageRank is more sensitive to find out the vertices in the outermost layer. (Color online.)}
				\label{directed_tree} 
			\end{figure}

			Now, we consider a directed trinary tree for our computation. Here, edges are directed outwards from a central vertex $0$. The vertices are distributed into four layers surrounding the central vertex. The vertices in the same layer have equal PageRank and qPageRank, which we mention in Table \ref{qPageRanks_for_directed_tree}. 
			\begin{table}[h!]
				\centering 
				\begin{tabular}{|c| c| c|}
					\hline 
					\textbf{Vertices in layers} & PageRank & qPageRank \\
					\hline
					Central vertex ($0$) & $0.005975$ & $0.0016$\\
					\hline
					Vertices in layer $1$ ($1, 2, 3$) & $0.007668 $ & $0.0020 $ \\
					\hline 
					Vertices in layer $2$ ($4, 5, \dots 12$) & $0.008148 $ & $0.0021$ \\
					\hline 
					Vertices in layer $3$ ($13, 14, \dots 39$) & $0.008283$ & $0.0022$\\
					\hline 
					Vertices in layer $4$ ($40, 41, \dots 120$ & $0.008322 $ & $0.0113$.\\
					\hline
				\end{tabular}
				\caption{PageRank and qPageRank values for the vertices in a directed trinary tree.}
				\label{qPageRanks_for_directed_tree}
			\end{table}
			We plot the values of PageRank and qPageRank of the directed tree using bar diagrams. The graph and its corresponding bar diagrams are kept in Figure \ref{directed_tree}. The PageRank and qPageRank indicate the same ranking pattern on the directed trinary tree. As a result, the rank correlation is $1$. Also, the difference in PageRank of the vertices in the third and fourth layers is $0.000039$. But, the difference in qPageRank of the vertices in the third and fourth layers is $.0091$. Therefore, the qPageRank is more sensitive for determining the vertices in the outer layer. Also, compare the bar diagrams in Figures \ref{balanced_tree_3_4_bar} and \ref{trinary_tree_bar}, which are depicted for the undirected and directed trinary trees with an equal number of vertices, respectively. The qPageRank is more efficient than the classical PageRank for finding the vertices in the outermost layer in the case of a directed trinary tree. 
			
\begin{figure}[h!]
				\begin{subfigure}[a]{.55\textwidth}
					\centering 
					\includegraphics[scale = .6]{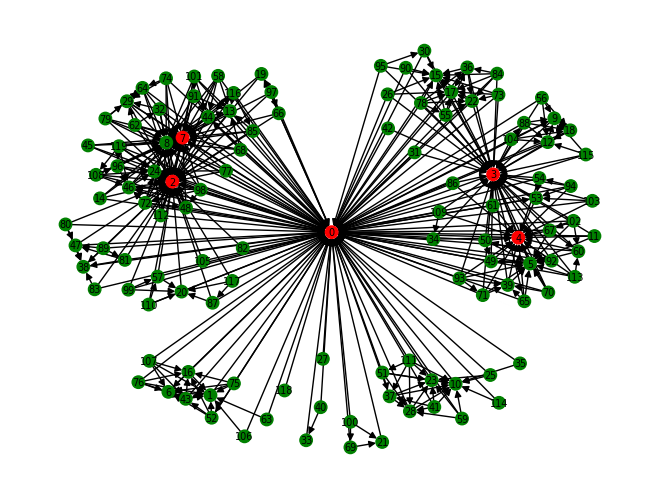}
					\caption{\textbf{GNC graph.} The vertices $0, 2, 3, 7$, and $4$ are in red color. They are the top five vertices with maximum PageRank and qPageRank.}
					\label{GNC_graph} 
				\end{subfigure}
				\hspace{.25cm}
				\begin{subfigure}[a]{.44\textwidth}
					\centering 
					\includegraphics[width = 6cm, height = 8cm]{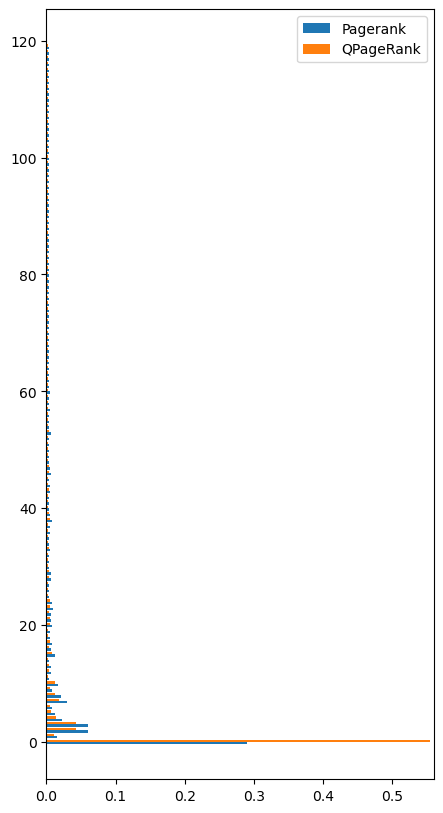}
					\caption{PageRank and qPageRank of the vertices of the graphs in figure \ref{GNC_graph} are represented with horizontal bar diagraph.} 
				\end{subfigure}
				\caption{A GNC graph with $120$ vertices and $478$ edges is depicted in figure \ref{GNC_graph}. The Kendall's rank correlation coefficient between the rank of the vertices generated by PageRank and qPageRank is $0.777311$. (Color online.)} 
			\end{figure}

			There are others directed random graphs for which PageRank and qPageRank behave similarly. The GNC graph serves as one example \cite{krapivsky2005network}. For our experiment, we consider a GNC graph with $120$ vertices and $478$ edges, which is depicted in figure \ref{GNC_graph}. The top five vertices with maximum PageRank and qPageRank are $0, 2, 3, 7$, and $4$. Kendall's rank correlation coefficient between the ranks of the verticse generated by PageRank and qPageRank is $0.777311$. The vertex $0$ has maximum PageRank and qPageRank. For qPageRank and PageRank, the values of vertex $0$ are $0.5545$ and $0.289865$, respectively. The qPageRank works more efficiently than the PageRank for detecting the node with maximum importance, for this graph. 
			
\begin{figure}[h!]
				\begin{subfigure}[a]{.55\textwidth}
					\centering 
					\includegraphics[scale = .6]{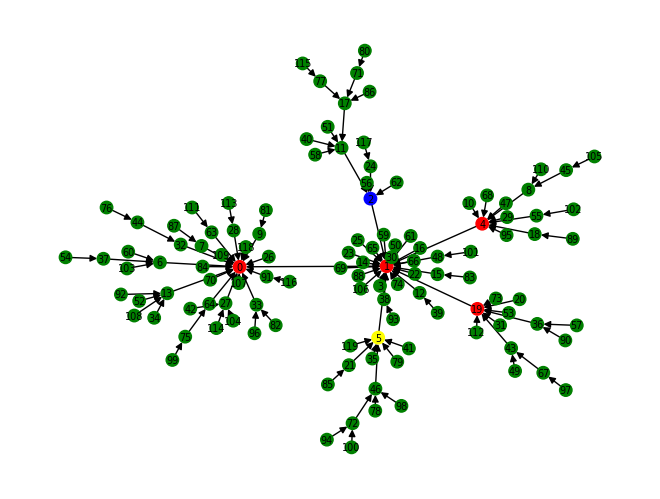}
					\caption{\textbf{GN graph.} The vertices $0, 1, 4, 19$ are in red color. They are among the top five vertices with maximum PageRank and qPageRank. The vertex $5$ is in yellow. It is among the top five with respect to qPageRank but not with respect to PageRank. Similarly, the blue vertex $2$ is in the top five vertices with respect to maximum PageRank but not with respect to maximum qPageRank.}
					\label{GN_graph} 
				\end{subfigure}
				\hspace{.25cm}
				\begin{subfigure}[a]{.44\textwidth}
					\centering 
					\includegraphics[width = 6cm, height = 8cm]{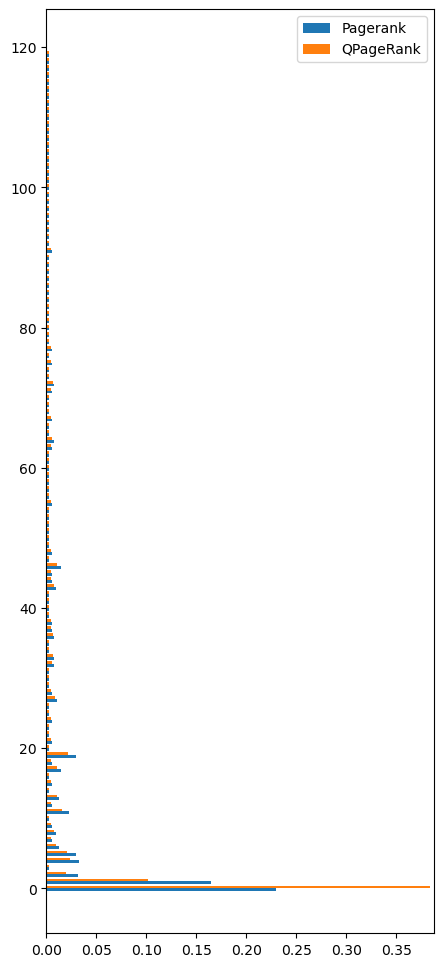}
					\caption{PageRank and qPageRank of the vertices of the graphs in figure \ref{GN_graph} are represented with horizontal bar diagraph.} 
				\end{subfigure}
				\caption{A GN graph with $120$ vertices and $119$ edges is depicted in figure \ref{GNC_graph}. The Kendall's rank correlation coefficient between the rank of the vertices generated by PageRank and qPageRank is $0.99944$. (Color online.)} 
			\end{figure}

			Next, we consider the GN graph \cite{krapivsky2001organization} with $120$ vertices and $119$ edges, which is depicted in figure \ref{GN_graph}. Kendall's rank correlation between the PageRank and qPageRank is superior to the GNC graph under our consideration. Kendall's rank correlation is $0.99944$. The top five vertices with maximum PageRank are $0, 1,4, 2$ and $19$ whereas the top five vertices with maximum qPageRank are $0, 1, 4, 19, 5$ and $15$. 
			
\begin{figure}[h!]
				\begin{subfigure}[a]{.55\textwidth}
					\centering 
					\includegraphics[scale = .6]{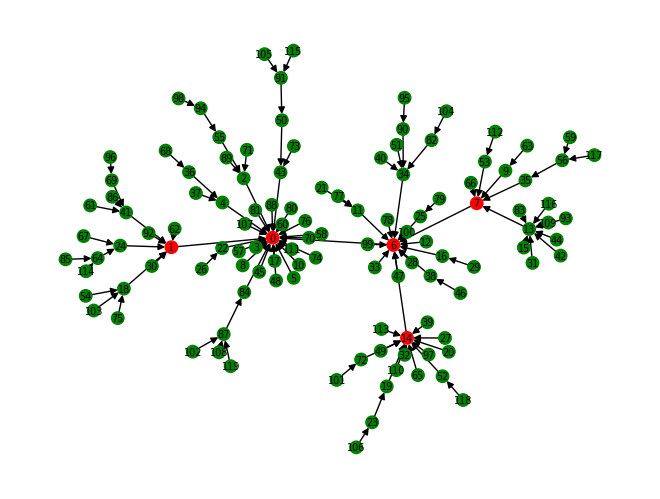}
					\caption{\textbf{GNR graph.} The top five vertices with respect to PageRank and qPageRank are red in color, which are $0, 6, 14, 7$, and $1$.}
					\label{GNR_graph} 
				\end{subfigure}
				\hspace{.25cm}
				\begin{subfigure}[a]{.44\textwidth}
					\centering 
					\includegraphics[width = 6cm, height = 7.5cm]{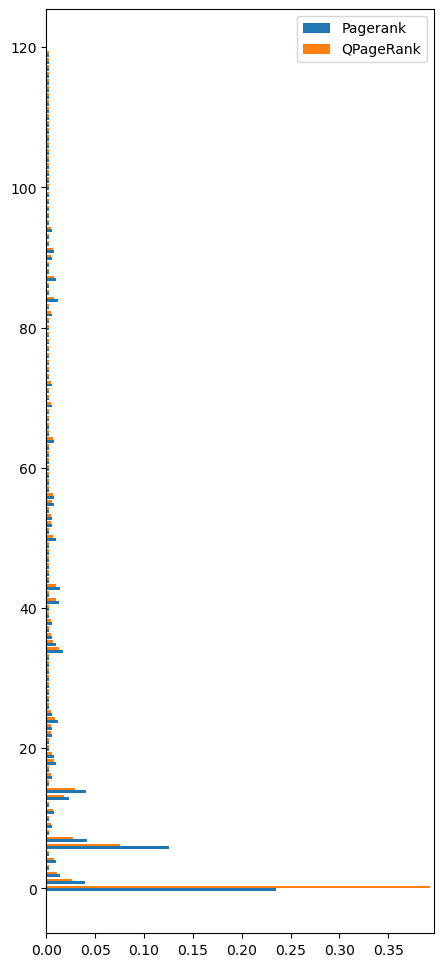}
					\caption{PageRank and qPageRank of the vertices of the graphs in figure \ref{GNR_graph} are represented with horizontal bar diagraph.} 
				\end{subfigure}
				\caption{A GNR graph with $120$ vertices and $119$ edges is depicted in figure \ref{GNR_graph}. The Kendall's rank correlation coefficient between the rank of the vertices generated by PageRank and qPageRank is $0.99916$. (Color online.)} 
			\end{figure}

			Next we consider the GNR graph \cite{krapivsky2001organization} with $120$ vertices and $119$ edges, which is depicted in figure \ref{GNR_graph}. In this graph, both PageRank and qPageRank exhibit similar behavior. The top five vertices with maximum PageRank are $0, 6, 7. 14$, and $1$. These vertices are also the top five vertices with maximum qPageRank, but in a different order, which is $0, 6, 14, 7$, and $1$.
			
			We also calculate PageRank and qPageRank for directed scale-free graph \cite{bollobas2003directed}, Erd\H{o}s-R\'enyi random graph \cite{erdos1960evolution}, random k-out graph \cite{peterson2015distance}. For the Erd\H{o}s-R\'enyi random graph and random $k$-out graph, there is a significant mismatch between PageRank and qPageRank of the vertices. For these directed graphs, the correlation coefficient between two rankings is much less than $1$. Note that, there is no trivial symmetry in these graphs \cite{Erdos1963asymmetric}. It has already been observed that the quantum PageRank and the classical PageRank do not always match \cite{paparo2013quantum, ortega2023generalized}. Our observation is also similar.
			
\begin{figure}[h!]
				\begin{subfigure}[a]{.65\textwidth}
					\centering
					\includegraphics[scale = .6]{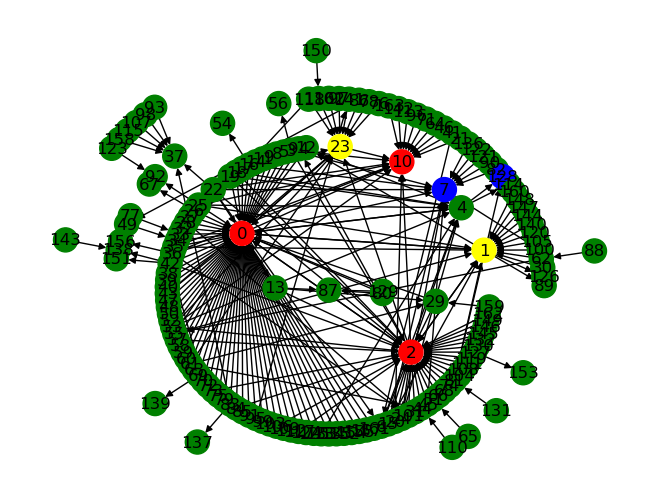}
					\caption{\textbf{A directed scale free graph}. The red vertices which are $0,  2$ and $10$ are among top five vertices with PageRank and qPageRank. The yellow vertices $1$ and $23$ are among the top five vertices with maximum PageRank but not in the top five vertices with maximum qPageRank. Similarly, the blue vertices $7$ and $128$ are in top five vertices with maximum qPageRank but not in the top five vertices with maximum PageRank.}
					\label{scale_free} 
				\end{subfigure}
				\hspace{.25cm}
				\begin{subfigure}[a]{.34\textwidth}
					\includegraphics[width = 6cm, height = 8cm]{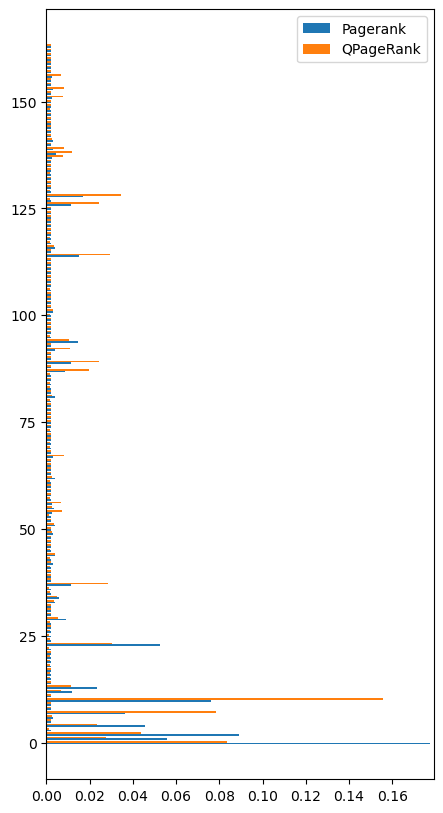}
					\caption{PageRank and qPageRank of the vertices of the scale free network in Figure \ref{scale_free} are plotted with horizontal bar diagram.}
				\end{subfigure}
				\caption{A directed scale free graph with $164$ vertices and $240$ edges. We find the ranks of the vertices based on PageRank and qPageRank. The Kendell's rank correlation coefficient is $0.607063$. (Color online.)}
			\end{figure}

			Consider the directed scale-free graph \cite{bollobas2003directed} with $164$ vertices and $240$ edges, which is depicted in Figure \ref{scale_free}. The top five vertices with maximum PageRank are $0, 2, 10, 1$, and $23$. Also, the top five vertices with maximum qPageRank are $10, 0, 7, 2$, and $128$. Both PageRank and qPageRank can identify the top three vertices. But, they arrange these vertices in a different order. The vertices $0, 2$ and $10$ are common in the top five vertices with the highest ranks. The qPageRank and PageRank behave differently. 
			
\begin{figure}[h!]
				\begin{subfigure}[a]{.65\textwidth}
					\centering
					\includegraphics[scale = .6]{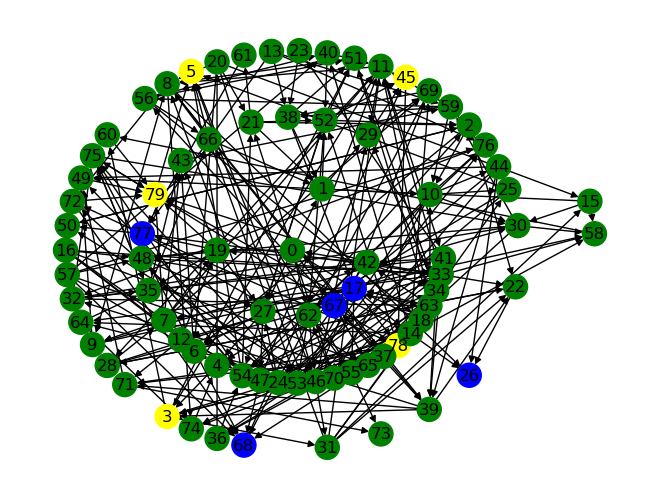}
					\caption{\textbf{A directed Erd\H{o}s-R\'enyi random graph.} The vertices $5, 79, 78, 45$, and $3$ are in yellow. They are the top five vertices with maximum PageRank. The vertices $26, 67, 77, 68$, and $17$ are in blue. They are the vertices with top qPageRank.} 
					\label{Erdos_Renyi} 
				\end{subfigure}
				\hspace{.25cm}
				\begin{subfigure}[a]{.34\textwidth}
					\includegraphics[width=6cm, height = 8cm]{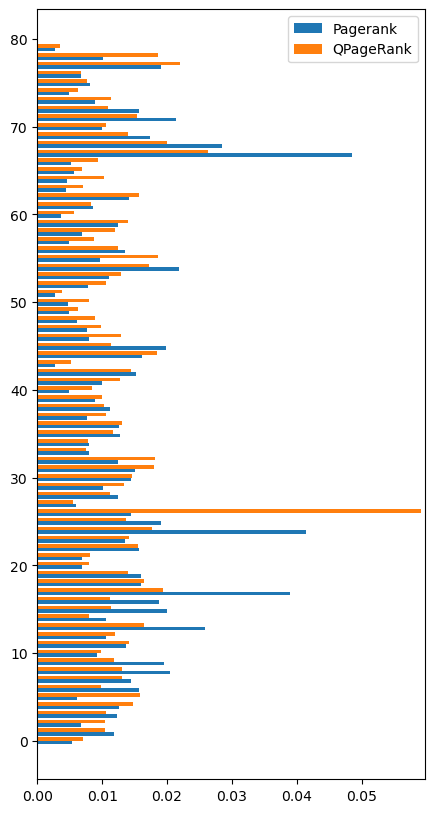}
					\caption{The PageRank and qPageRank of the vertices in the Erd\H{o}s-R\'enyi random graph of Figure \ref{Erdos_Renyi} are depicted with horizental bars.}
				\end{subfigure}
				\caption{Erd\H{o}s-R\'enyi random graph with $80$ vertices and $283$ edges. We consider the probability of existence of an edge is $p = 0.1$. We find the ranks of the vertices based on PageRank and qPageRank. The Kendell's rank correlation coefficient is $0.003165$. (Color online.)}
			\end{figure}

			The Erd\H{o}s-R\'enyi random graph \cite{erdos1960evolution} is a family of well-known random graphs. The edges are selected randomly to construct a random graph. We consider a specific instance of a directed Erd\H{o}s-R\'enyi random graph with $80$ vertices and $283$ edges, which we plot in figure \ref{Erdos_Renyi}. In this case, the PageRank and qPageRank behave differently. The top five vertices with the highest PageRank are $5, 79, 78, 45$, and $3$. Also, the top five vertices with the highest qPageRank are $26, 67, 77, 68$, and $17$. There is no common vertex. The difference between PageRank and qPageRank for the Erd\H{o}s-R\'enyi random graph is also talked about in other qPageRank proposals \cite{paparo2013quantum, ortega2023generalized}.
			
\begin{figure}[h!]
				\begin{subfigure}[a]{.55\textwidth}
					\centering 
					\includegraphics[scale = .6]{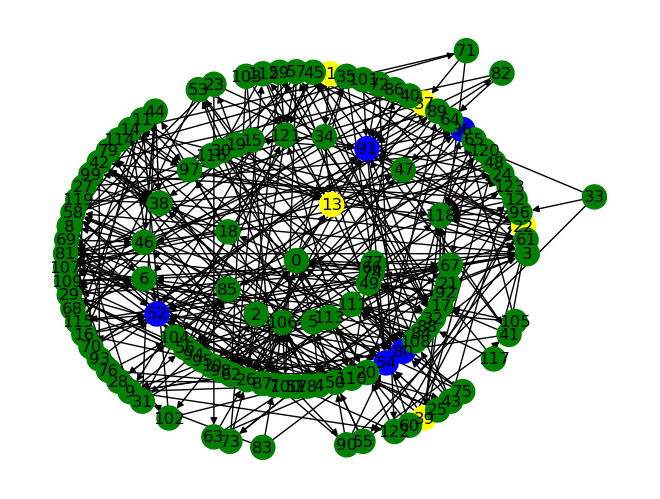}
					\caption{\textbf{A directed random $k$-out graph.} The top five vertices with highest PageRank are in yellow color. Also, top five vertices with highest qPageRank are in blue color.}
					\label{Random_k_out_graph} 
				\end{subfigure}
				\hspace{.25cm}
				\begin{subfigure}[a]{.44\textwidth}
					\centering 
					\includegraphics[width=6cm, height = 8cm]{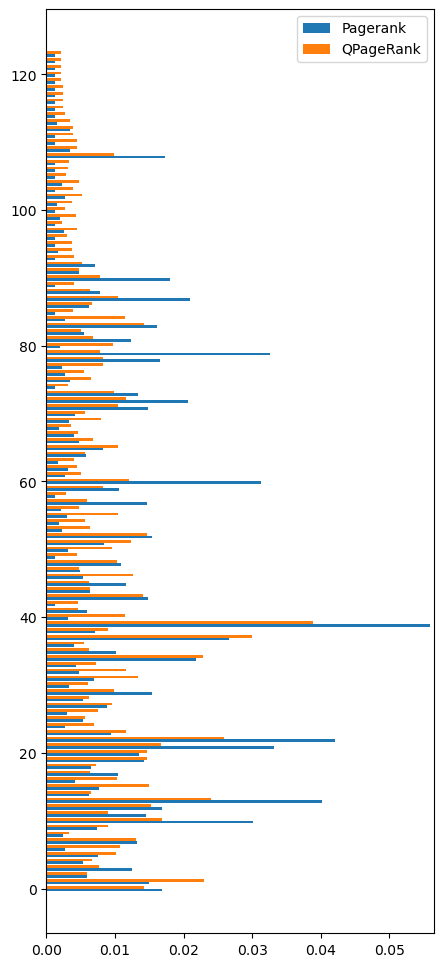}
					\caption{PageRanks and qPageRanks of the vertices of the graph in figure \ref{Random_k_out_graph} are represented by horizontal blue and red bars, respectively.} 
				\end{subfigure}
				\caption{A random $k$ out graph with $124$ vertices and $371$ edges is depicted in figure \ref{Random_k_out_graph}. The Kendall's rank correlation coefficient is $0.033307$. (Color online.)} 
			\end{figure}

			Other than the Erd\H{o}s-R\'enyi random graph, there are graphs for which the PageRank and qPageRank behave differently. One such graph is the random $k$-out graph \cite{peterson2015distance}. A specific instance of this graph with $124$ vertices and $371$ edges is depicted in Figure \ref{Random_k_out_graph}. The top five vertices with maximum PageRank are $91, 52, 54, 80$, and $36$. Also, the top five vertices with maximum qPageRank are $39, 37, 22, 13$, and $1$. We rank the vertices according to PageRank and qPageRank. The Kendall's rank correlation coefficient is $0.033307$. 
			
		\subsection{Convergence test of qPageRank}
		\label{Convergence_test_of_qPageRank} 
		
			In this section, we illustrate the convergence of the sequence of qPageRank $q\Pi^{(t)} = (p_1^{(t)}, p_2^{(t)}, \dots, p_n^{(t)})$ for $t = 0, 1, \dots \infty$ on different graphs. Recall that, we have mentioned the definition of qPageRank in Definition \ref{qPageRank_definiiton}. Also, the stopping criterion for determining the convergence of qPageRank procedure is mentioned in equation (\ref{stopping_creterion}). Initially, we consider that the walker can start walking from any vertex chosen randomly and uniformly. The initial state is described in equation (\ref{initial_state}). It indicates that every vertex initially has equal qPageRank. Every step of the quantum walk updates the qPageRank value of a vertex until it reaches a limiting value.
			
\begin{figure}
				\centering 
				\begin{subfigure}[a]{.48\textwidth}
					\centering 
					\includegraphics[height = 3cm, width = 8cm]{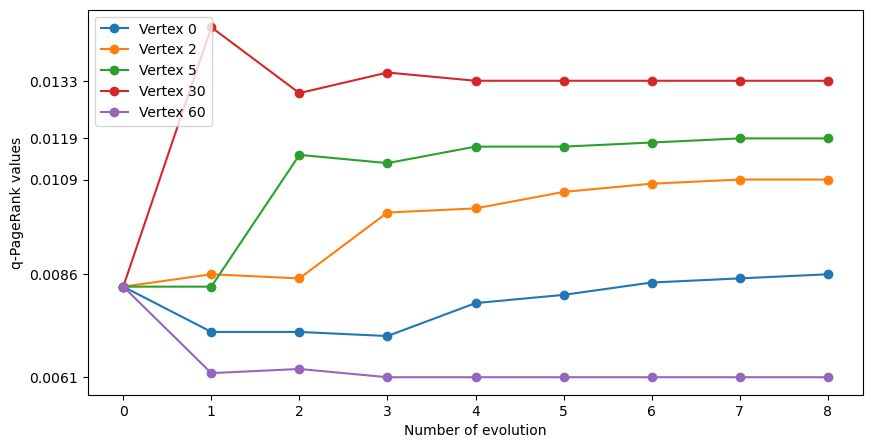}
					\caption{Convergence of qPageRank values in the \textbf{balanced tree $(3, 4)$} with $121$ vertices and $120$ edges.}
				\end{subfigure}
				\hspace{.3cm}
				\begin{subfigure}[a]{.48\textwidth}
					\centering 
					\includegraphics[height = 3cm, width = 8cm]{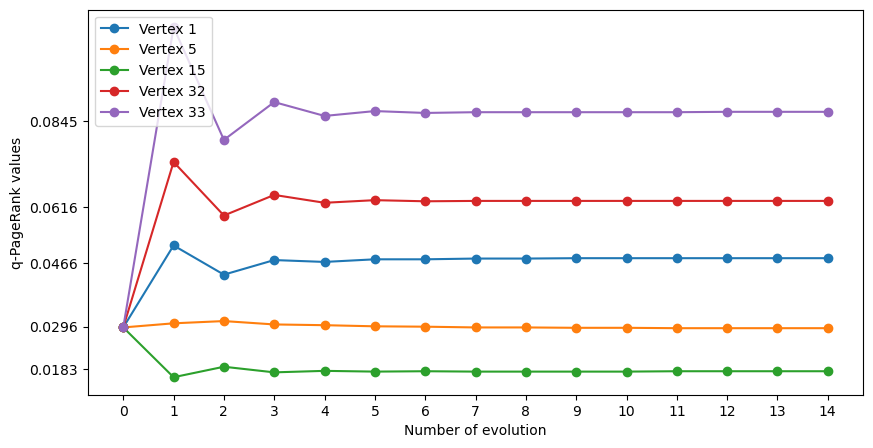}
					\caption{Convergence of qPageRank values in the \textbf{Karate Club} network with $34$ vertices and $78$ edges.}
				\end{subfigure}
				\\
				\begin{subfigure}[a]{.48\textwidth}
					\centering 
					\includegraphics[height = 3cm, width = 8cm]{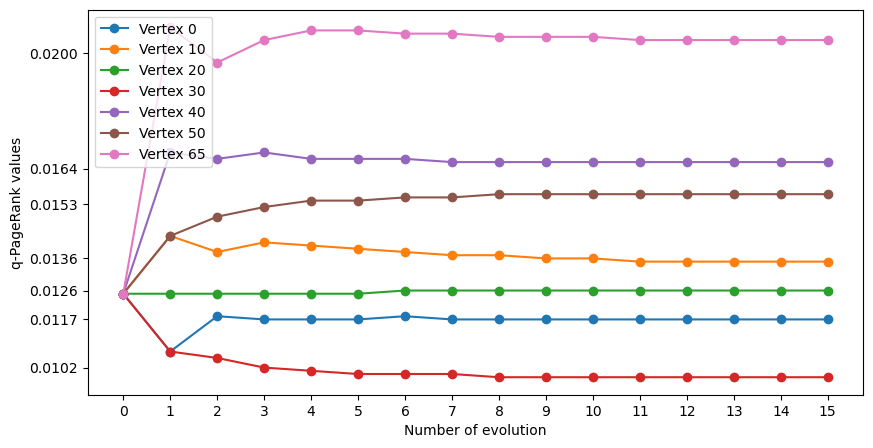}
					\caption{Convergence of qPageRank values in the \textbf{Watts-Strogatz network} with $80$ vertices and $80$ edges.}
				\end{subfigure}
				\hspace{.3cm}
				\begin{subfigure}[a]{.48\textwidth}
					\centering 
					\includegraphics[height = 3cm, width = 7.5cm]{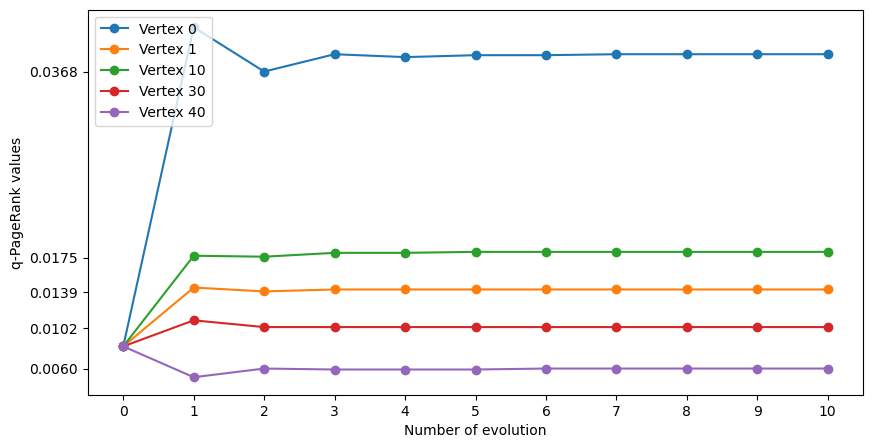}
					\caption{Convergence of qPageRank values in the \textbf{Barabási and Albert network} with $120$ vertices and $464$ edges.}
				\end{subfigure}
				\\
				\begin{subfigure}[a]{.48\textwidth}
					\centering 
					\includegraphics[height = 3cm, width = 7.5cm]{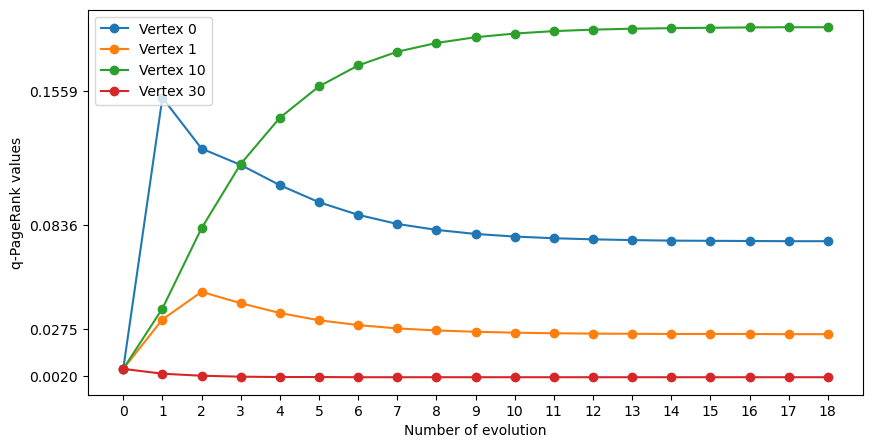}
					\caption{Convergence of qPageRank values in a \textbf{Scale free graph} with $164$ vertices and $240$ edges.}
				\end{subfigure}
				\hspace{.3cm}
				\begin{subfigure}[a]{.48\textwidth}
					\centering 
					\includegraphics[height = 3cm, width = 7.5cm]{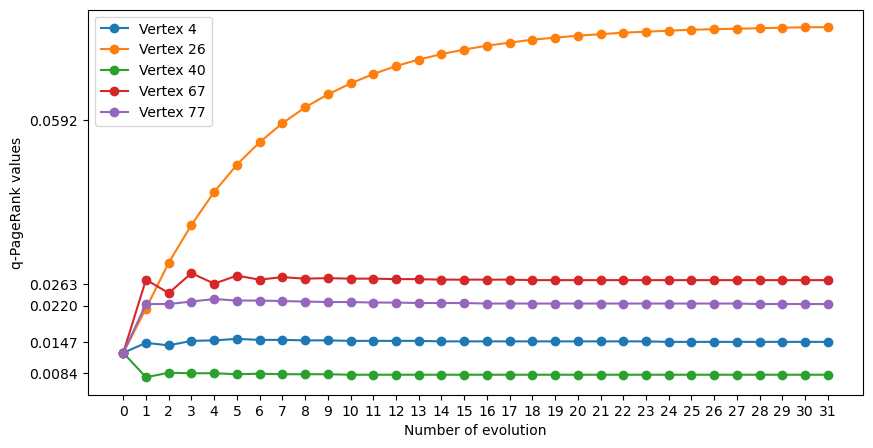}
					\caption{Convergence of qPageRank values in a \textbf{Erdos-Renyi random network} with $80$ vertices and $283$ edges.}
				\end{subfigure}
				\\
				\begin{subfigure}[a]{.48\textwidth}
					\centering 
					\includegraphics[height = 3cm, width = 7.5cm]{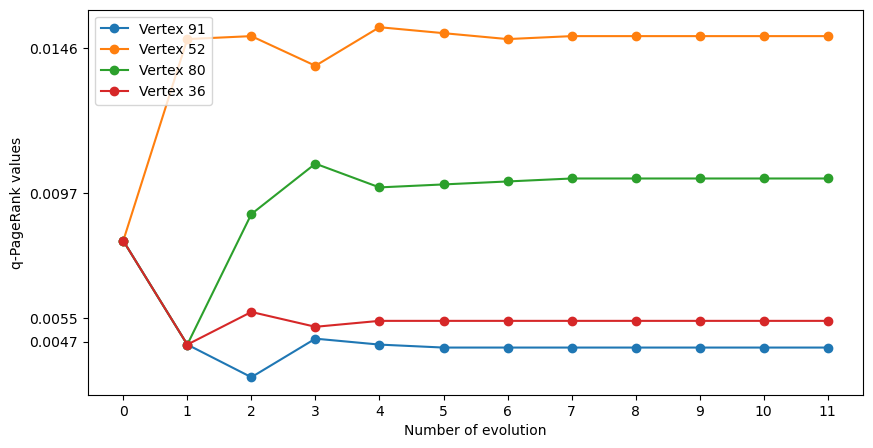}
					\caption{Convergence of qPageRank values in a \textbf{Random $k$-out graph} with $124$ vertices and $371$ edges.}
				\end{subfigure}
				\hspace{.3cm}
				\begin{subfigure}[a]{.48\textwidth}
					\centering 
					\includegraphics[height = 3cm, width = 7.5cm]{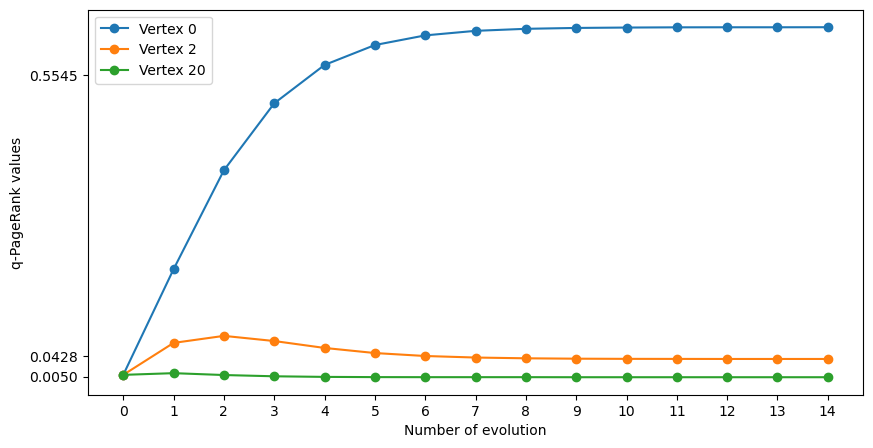}
					\caption{Convergence of qPageRank values in a \textbf{GNC graph} with $120$ vertices and $478$ edges.}
				\end{subfigure}
				\\
				\begin{subfigure}[a]{.48\textwidth}
					\centering 
					\includegraphics[height = 3cm, width = 7.5cm]{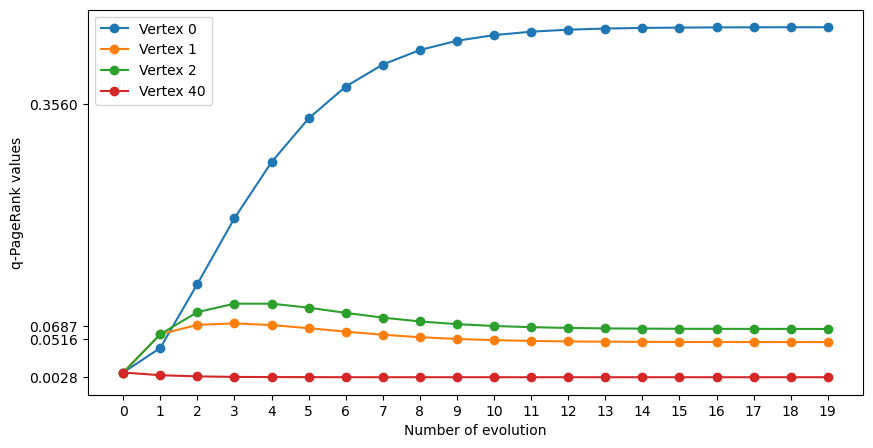}
					\caption{Convergence of qPageRank values in a \textbf{GN graph} with $120$ vertices and $119$ edges.}
				\end{subfigure}
				\hspace{.3cm}
				\begin{subfigure}[a]{.48\textwidth}
					\centering 
					\includegraphics[height = 3cm, width = 7.5cm]{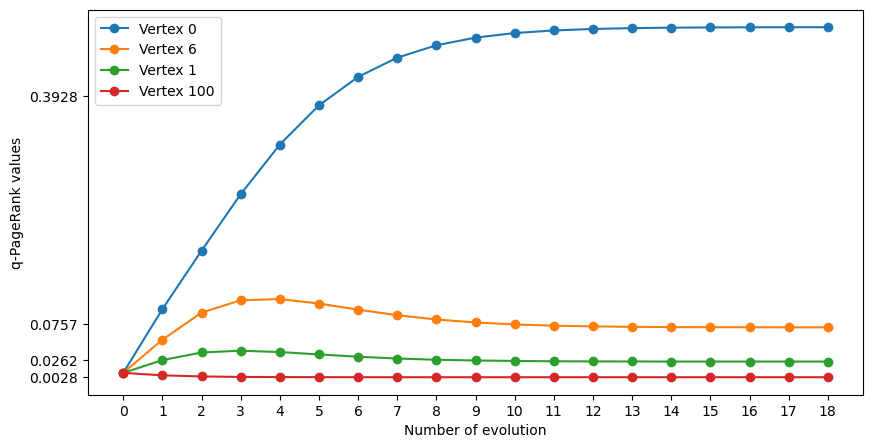}
					\caption{Convergence of qPageRank values in a \textbf{GNR graph} with $120$ vertices and $119$ edges.}
				\end{subfigure}
				\caption{In the sub-figures we plot the qPageRank values at different step of quantum walk for a few vertices in different graphs. We keep the number of steps of quantum walk in $x$-axis. In $y$-axis we keep the qPageRank values at different time steps. Given any vertex we join the consecutive values of qPageRanks by a line segment. It indicates the convergence of qPageRank for the vertices in the sample. (Color online.)}
				\label{convergence_tests}
			\end{figure}

			For our numerical experiment to determine convergence of qPageRank procedure, we consider a random sample of the vertices from a graph, which includes the vertex with the maximum qPageRank. For different time instance $t = 0, 1, 2, \dots T$, we calculate $q\Pi^{(t)}$ and plot corresponding values for distinct vertices in our sample. Note that, we consider accuracy of $q\Pi$ upto three decimal places. Therefore, $q\Pi^{(T)}$ represents the accurate value of $q\Pi$ upto three decimal places. For better accuracy the number of iterations may be increased. For different graphs we collect the plots in Figure \ref{convergence_tests}.
			
			We observe that the number of evolutions for convergence does not necessarily depend on the size of the graphs. We need thirty-one steps of quantum walk to achieve the convergence of the qPageRank sequence of the Erdos-Renyi random graph with $80$ vertices and $283$ edges. It is the maximum among all the graphs considered in this article. In contrast, for the GNC graph with $120$ vertices and $478$ edges, the qPageRank sequence takes only $14$ steps of evolution to reach the limiting value.
			
			Similar numerical investigations on the convergence of qPagerank are reported in \cite{loke2017comparing, chawla2020discrete, ortega2023generalized}. In \cite{chawla2020discrete} convergence is observed in more than $50$ steps. Article \cite{ortega2023generalized} uses several thousands of iterations of quantum walk for studying convergence. Our method takes a smaller number of iterations for finding the qPageRank than any of them.

		\subsection{Dependency of qPageRank on $\alpha$}
		
			There are experimental evidences showing that the PageRank exhibits a dependency on the damping parameter $\alpha$. In fact, the choice of $\alpha$ in the classical PageRank algorithm is also motivated by the need to ensure fast convergence of the power method. It ensures a reasonably large spectral gap in the Google matrix. The usual choice of $\alpha = 0.85$ is expected to produce a satisfactory web surfing model, as well as a good convergence rate in computing the PageRank vector. The stability of the qPageRank is also a crucial factor to consider, as it randomly sets the damping parameter to a certain value. In fact, there is no a priori argument available to set the value of $\alpha$. The initial value of $0.85$ in the conventional PageRank protocol was selected to resemble the actions of a random walker. Given the arbitrary selection of the parameter $\alpha$, it is highly preferred that the algorithm's output remain stable. Hence, it is interesting to investigate how the qPageRank is inspired by the choice of $\alpha$. For the other proposals, the dependency of qPageRank on $\alpha$ is investigated in \cite{paparo2012google, paparo2013quantum, paparo2014quantum, ortega2023generalized, boito2023ranking}
			
			To understand the dependency of $\alpha$ on qPageRank, we work out an average state and a final state for every value of $\alpha$. Recall that in equation (\ref{evolution_with_damping_factor}), we describe the state $\rho^{(t)}$ for a given value $\alpha$ at a particular time instance. The terminating criterion of Procedure \ref{qPageRank_algo} is mentioned in equation (\ref{stopping_creterion}). Let we terminate the process at time $T$. Then the final state is $\rho^{(T)}$ and the average state is 
			\begin{equation}\label{average_state}
				\rho_{average} = \frac{1}{T + 1} \sum_{t = 0}^T \rho^{(t)}.
			\end{equation} 
			
\begin{figure}	
				\centering 
				\begin{subfigure}[a]{.48\textwidth}
					\centering 
					\includegraphics[height = 4cm, width = 8cm]{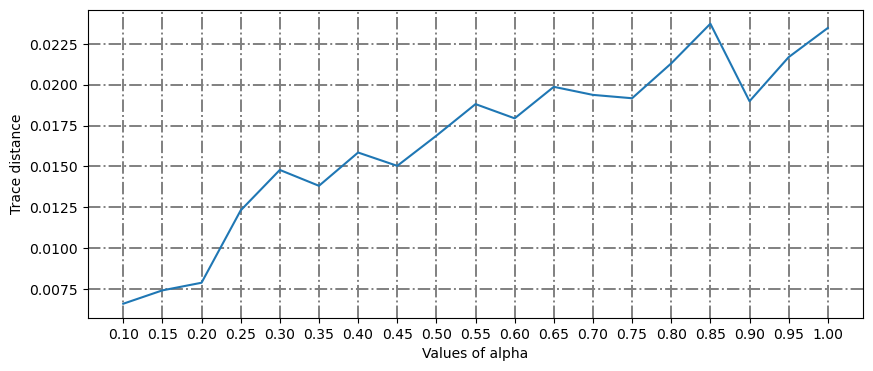}
					\caption{Trace distance between the average state and final state with respect to $\alpha$ for a \textbf{balanced tree $(3, 4)$} with $121$ vertices and $120$ edges.}
					\label{Balanced_tree_alpha_trace_distance} 
				\end{subfigure}
				\hspace{.3cm}
				\begin{subfigure}[a]{.48\textwidth}
					\centering 
					\includegraphics[height = 4cm, width = 8cm]{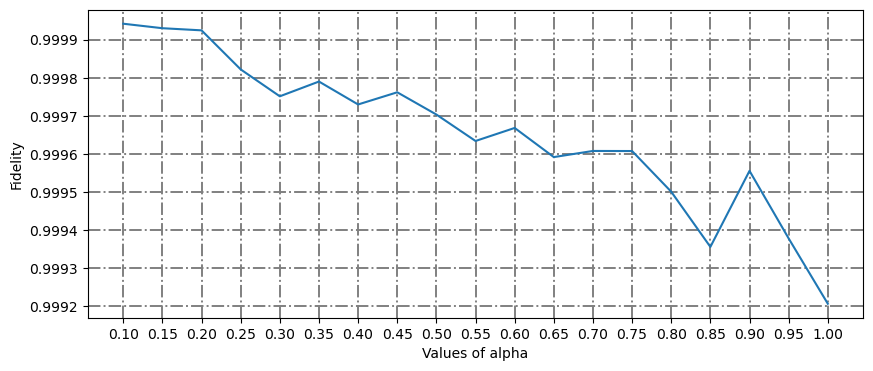}
					\caption{Fidelity between the average state and final state with respect to $\alpha$ for a \textbf{balanced tree $(3, 4)$} with $121$ vertices and $120$ edges.}
					\label{Balanced_tree_alpha_fidelity} 
				\end{subfigure}
				\\
				\begin{subfigure}[a]{.48\textwidth}
					\centering 
					\includegraphics[height = 4cm, width = 8cm]{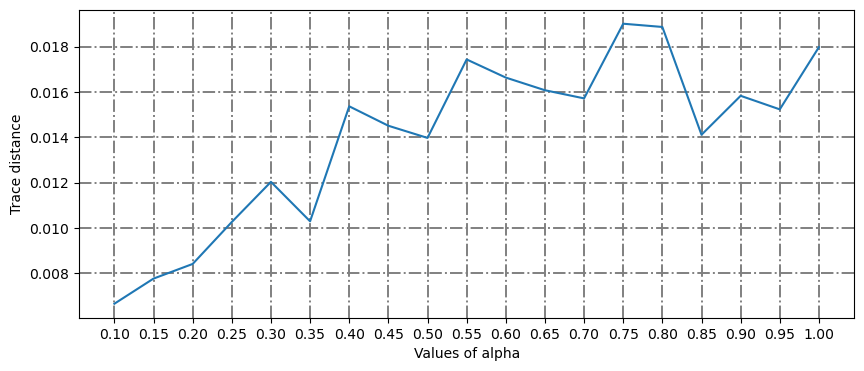}
					\caption{Trace distance between the average state and final state with respect to $\alpha$ for the \textbf{Karate Club} network.}
					\label{Karate_club_alpha_trace_distance}
				\end{subfigure}
				\hspace{.3cm}
				\begin{subfigure}[a]{.48\textwidth}
					\centering 
					\includegraphics[height = 4cm, width = 8cm]{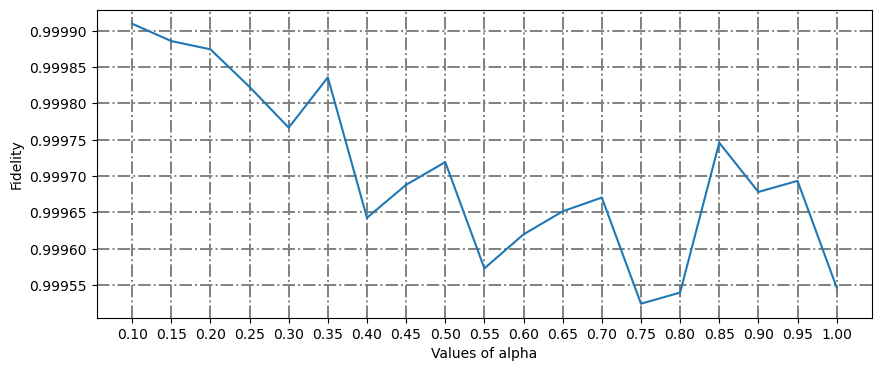}
					\caption{Fidelity between the average state and final state with respect to $\alpha$ for the \textbf{Karate Club} network.}
					\label{Karate_club_alpha_fidelity} 
				\end{subfigure}
				\\
				\begin{subfigure}[a]{.48\textwidth}
					\centering 
					\includegraphics[height = 4cm, width = 8cm]{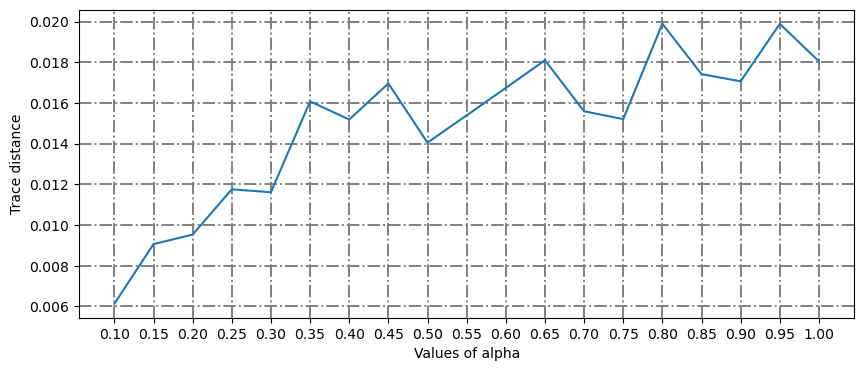}
					\caption{Trace distance between the average state and final state with respect to $\alpha$ for a \textbf{Barabási and Albert network}.}
					\label{BA_alpha_trace_distance} 
				\end{subfigure}
				\hspace{.3cm}
				\begin{subfigure}[a]{.48\textwidth}
					\centering 
					\includegraphics[height = 4cm, width = 8cm]{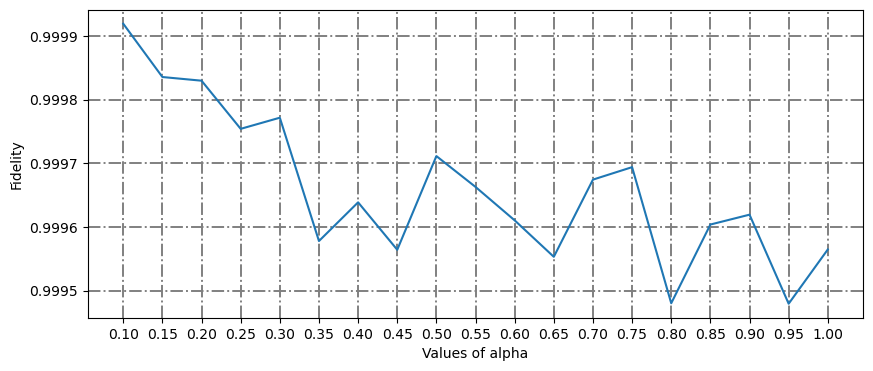}
					\caption{Fidelity between the average state and final state with respect to $\alpha$ for a \textbf{Barabási and Albert network}.}
					\label{BA_alpha_fidelity}
				\end{subfigure}	
				\\
				\begin{subfigure}[a]{.48\textwidth}
					\centering 
					\includegraphics[height = 4cm, width = 8cm]{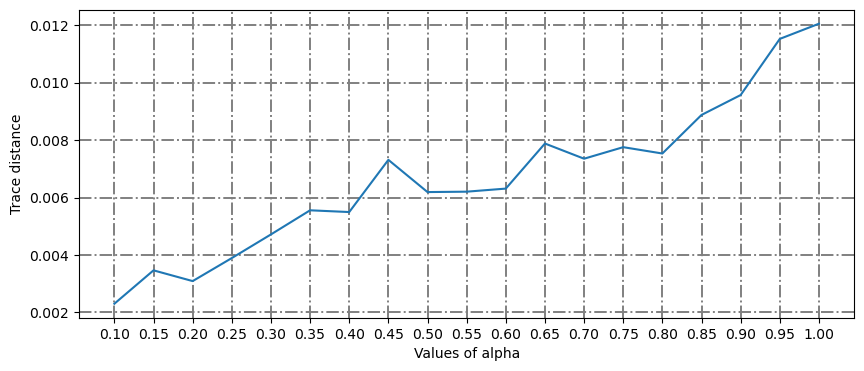}
					\caption{Trace distance between the average state and final state with respect to $\alpha$ for a \textbf{Watts-Strogatz network}.}
					\label{WS_alpha_trace_distance} 
				\end{subfigure}
				\hspace{.3cm}
				\begin{subfigure}[a]{.48\textwidth}
					\centering 
					\includegraphics[height = 4cm, width = 8cm]{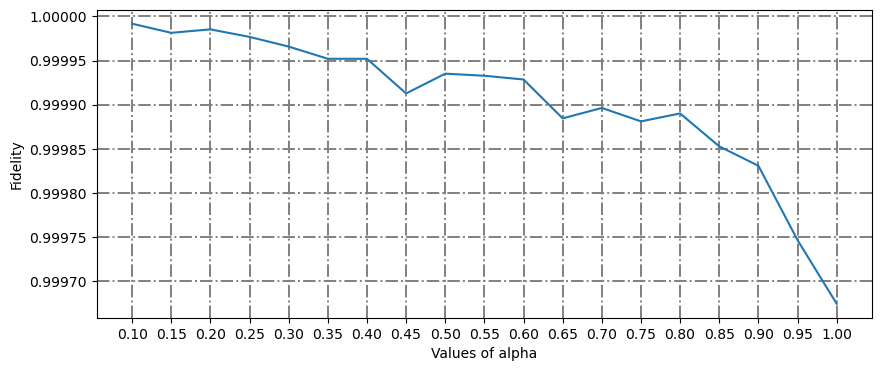}
					\caption{Fidelity between the average state and final state with respect to $\alpha$ for a \textbf{Watts-Strogatz network}.}
					\label{WS_alpha_fidelity} 
				\end{subfigure}
				\caption{In the sub-figures we plot the trace distance and fidelity between average state and final state with respect to $\alpha$ for a number of undirected graphs. (Color online.)}
				\label{alpha_undirected}
			\end{figure}

\begin{figure}
				\centering 
				\begin{subfigure}[a]{.48\textwidth}
					\centering 
					\includegraphics[height = 2.6cm, width = 8cm]{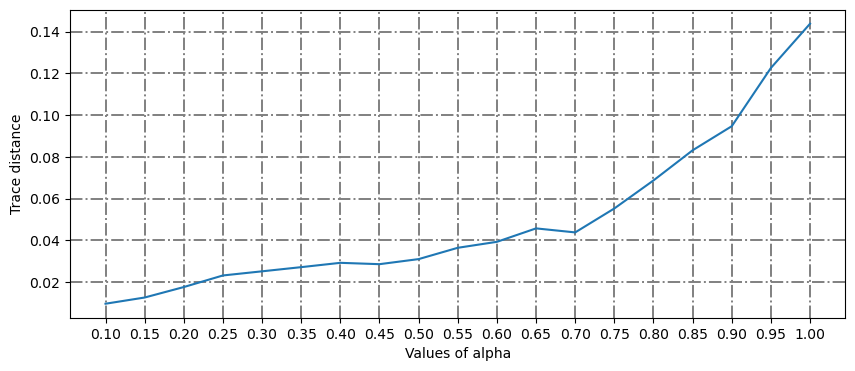}
					\caption{Trace distance between the average state and final state with respect to $\alpha$ for a \textbf{Scale-free} network.}
				\end{subfigure}
				\hspace{.3cm}
				\begin{subfigure}[a]{.48\textwidth}
					\centering 
					\includegraphics[height = 2.6cm, width = 8cm]{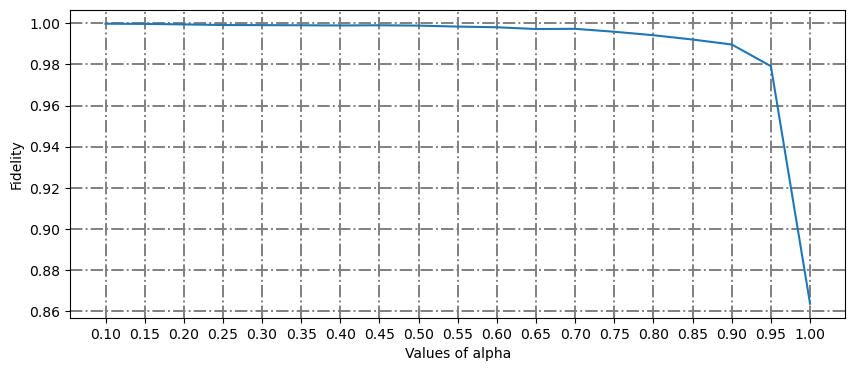}
					\caption{Fidelity between the average state and final state with respect to $\alpha$ for a \textbf{Scale free} network.}
				\end{subfigure}
				\\
				\begin{subfigure}[a]{.48\textwidth}
					\centering 
					\includegraphics[height = 2.6cm, width = 8cm]{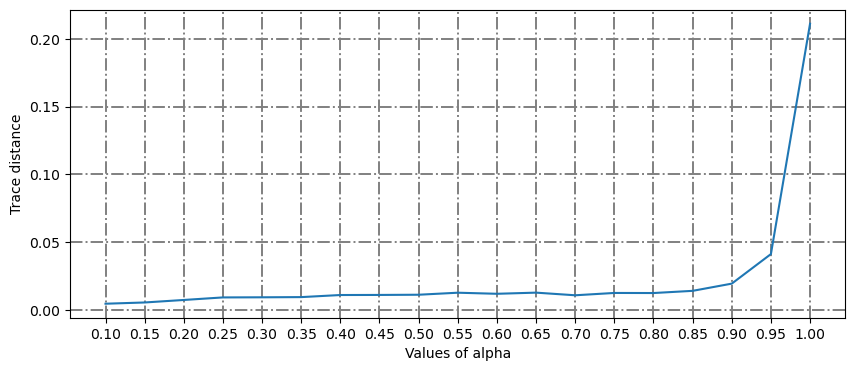}
					\caption{Trace distance between the average state and final state with respect to $\alpha$ for a \textbf{Erdos-Renyi random network}.}
				\end{subfigure}
				\hspace{.3cm}
				\begin{subfigure}[a]{.48\textwidth}
					\centering 
					\includegraphics[height = 2.6cm, width = 8cm]{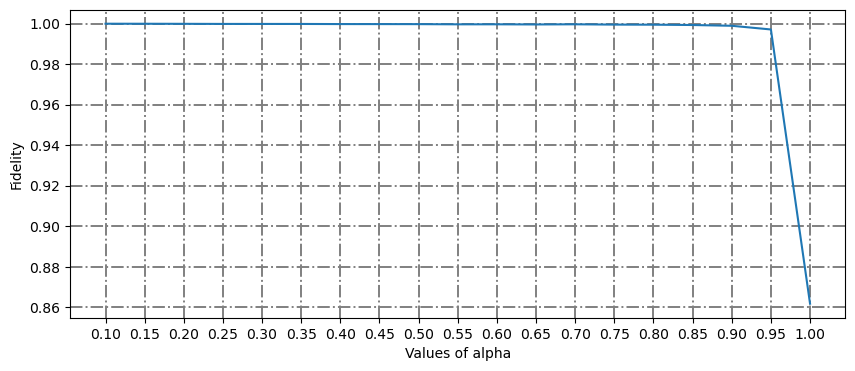}
					\caption{Fidelity between the average state and final state with respect to $\alpha$ for a \textbf{Erdos-Renyi random network}.}
				\end{subfigure}
				\\
				\begin{subfigure}[a]{.48\textwidth}
					\centering 
					\includegraphics[height = 2.6cm, width = 8cm]{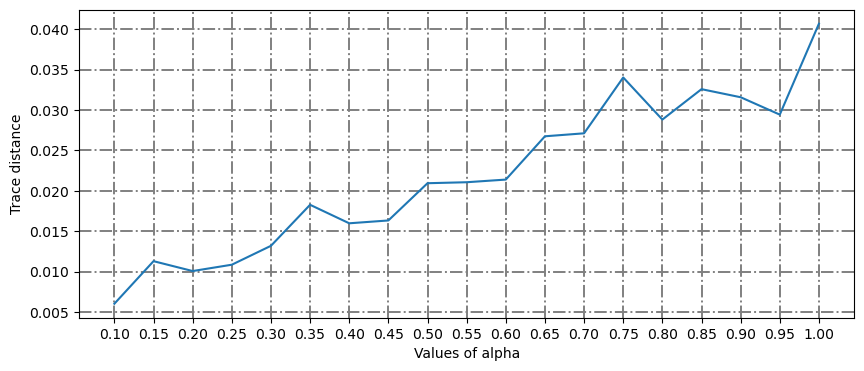}
					\caption{Trace distance between the average state and final state with respect to $\alpha$ for a \textbf{Random $k$-out graph}.}
				\end{subfigure}
				\hspace{.3cm}
				\begin{subfigure}[a]{.48\textwidth}
					\centering 
					\includegraphics[height = 2.6cm, width = 8cm]{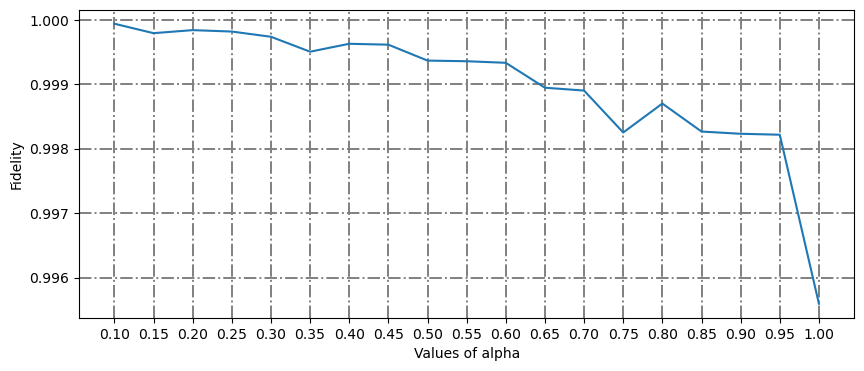}
					\caption{Fidelity between the average state and final state with respect to $\alpha$ for a \textbf{Random $k$-out graph}.}
				\end{subfigure}
				\\
				\begin{subfigure}[a]{.48\textwidth}
					\centering 
					\includegraphics[height = 2.6cm, width = 8cm]{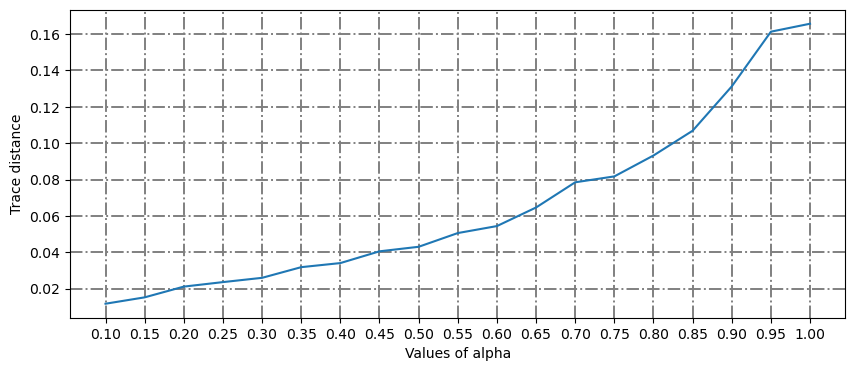}
					\caption{Trace distance between the average state and final state with respect to $\alpha$ for a \textbf{GNC graph}.}
				\end{subfigure}
				\hspace{.3cm}
				\begin{subfigure}[a]{.48\textwidth}
					\centering 
					\includegraphics[height = 2.6cm, width = 8cm]{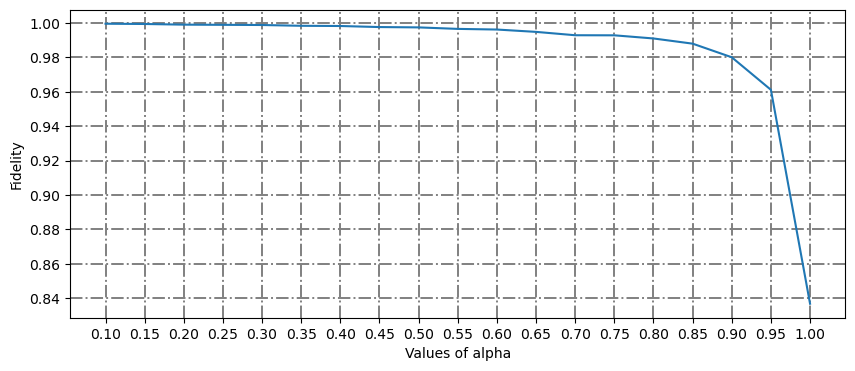}
					\caption{Fidelity between the average state and final state with respect to $\alpha$ for a \textbf{GNC graph}.}
				\end{subfigure}
				\\
				\begin{subfigure}[a]{.48\textwidth}
					\centering 
					\includegraphics[height = 2.6cm, width = 8cm]{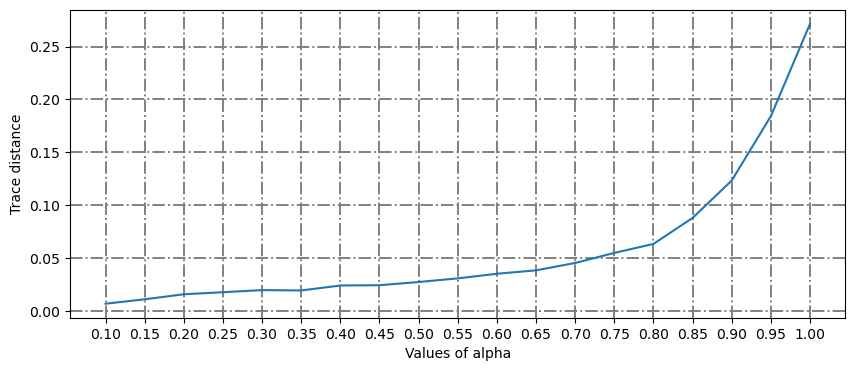}
					\caption{Trace distance between the average state and final state with respect to $\alpha$ for a \textbf{GN graph}.}
				\end{subfigure}
				\hspace{.3cm}
				\begin{subfigure}[a]{.48\textwidth}
					\centering 
					\includegraphics[height = 2.6cm, width = 8cm]{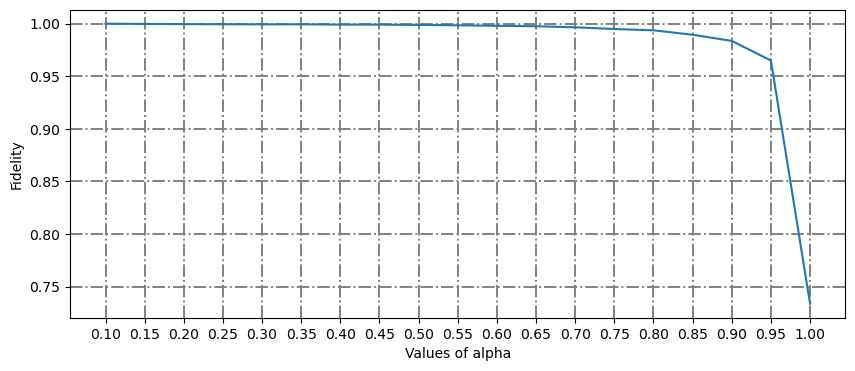}
					\caption{Fidelity between the average state and final state with respect to $\alpha$ for a \textbf{GN graph}.}
				\end{subfigure}
				\\
				\begin{subfigure}[a]{.48\textwidth}
					\centering 
					\includegraphics[height = 2.6cm, width = 8cm]{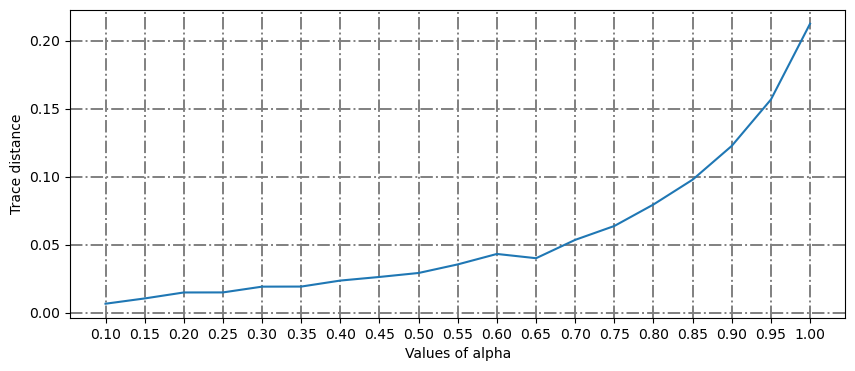}
					\caption{Trace distance between the average state and final state with respect to $\alpha$ for a \textbf{GNR graph}.}
				\end{subfigure}
				\hspace{.3cm}
				\begin{subfigure}[a]{.48\textwidth}
					\centering 
					\includegraphics[height = 2.6cm, width = 8cm]{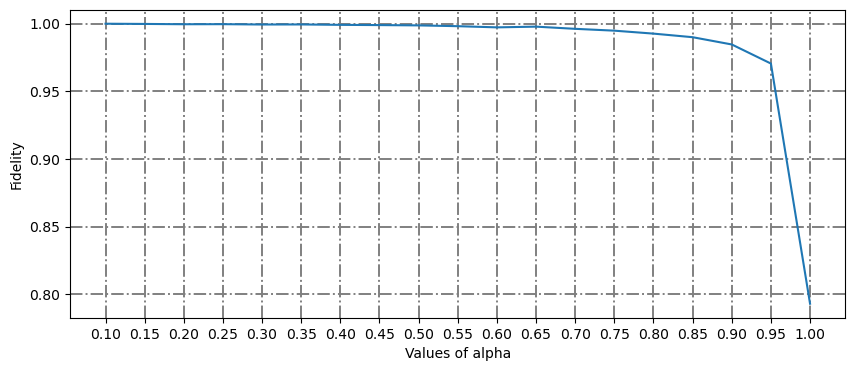}
					\caption{Fidelity between the average state and final state with respect to $\alpha$ for a \textbf{GNR graph}.}
				\end{subfigure}
				\caption{In the sub-figures we plot the trace distance and fidelity between average state and final state with respect to $\alpha$ for a number of directed graphs. (Color online.)}
				\label{alpha_directed}
			\end{figure}

			In this investigation, we use quantum fidelity and trace distance between quantum states to quantify their difference. The quantum fidelity between two quantum state represented by the density matrices $\rho$ and $\sigma$ is defined by
			\begin{equation}
				F(\sigma, \rho) = \left(\trace\left(\sqrt{\rho^{\frac{1}{2}} \sigma \rho^{\frac{1}{2}}}\right) \right)^2.
			\end{equation} 
			Also, the trace distance between $\rho$ and $\sigma$ is given by
			\begin{equation}
				D(\sigma, \rho) = \frac{1}{2}\trace |\rho - \sigma|.
			\end{equation} 
			We calculate the trace distance $D(\rho_{average}, \rho^{(T)})$ and fidelity $F(\rho_{average}, \rho^{(T)})$ between the average state $\rho_{average}$ and the final state $\rho^{(T)}$ for different values of $\alpha$. In figure \ref{alpha_undirected} and \ref{alpha_directed}, we plot $D(\rho_{average}, \rho^{(T)})$ and $F(\rho_{average}, \rho^{(T)})$ with respect to $\alpha$ for different undirected and directed graphs, respectively, under our consideration. Below we summaries our observations from these figures.
			
			Consider the undirected graphs first. We observe that if the trace distance $D(\rho_{average}, \rho^{(T)})$ has a local maxima for some $\alpha = \alpha_0 \in (0, 1)$, then the fidelity $F(\rho_{average}, \rho^{(T)})$ has a local minima at $\alpha_0$. For instance, consider the graph for $D(\rho_{average}, \rho^{(T)})$ and $F(\rho_{average}, \rho^{(T)})$, which are depicted in sub-figures \ref{Balanced_tree_alpha_trace_distance} and \ref{Balanced_tree_alpha_fidelity}, respectively, for the balanced tree $(3, 4)$. In sub-figure \ref{Balanced_tree_alpha_trace_distance} we observe a local maxima at $\alpha = 0.85$. Also, in sub-figure \ref{Balanced_tree_alpha_fidelity} we obtain a local minima at $\alpha = 0.85$. Similarly, these local maxima and minima can be observed at $\alpha = 0.75$ in the case of Karate club network, which are depicted in figures \ref{Karate_club_alpha_trace_distance} and \ref{Karate_club_alpha_fidelity}. The Barabási and Albert network has local maxima and minima for $\alpha = 0.95$, which is depicted in sub-figures \ref{BA_alpha_trace_distance} and \ref{BA_alpha_fidelity}. In the case of the Watts-Strogatz network there are no such local maxima or minima. The corresponding curves are drawn in sub-figure \ref{WS_alpha_trace_distance} and \ref{WS_alpha_fidelity}. Therefore, we may assume that $\alpha = 0.85$ is a wise choice. 
			
			We depict the curves for trace-distance and fidelity for directed graphs in figure \ref{alpha_directed}. The value $\alpha = 0.95$ has a special significance for directed graphs, under our consideration. The fidelity curves $F(\rho_{average}, \rho^{(T)})$ with respect to $\alpha$ sharply decline from $\alpha = .95$ for scale-free network, Erdos-Renyi random network, random $k$-out graph, GNC graph, GN graph, as well as GNR graph. The curve of trace-distance for Erdos-Renyi random network, random $k$-out graph and GNC graph increase sharply from $\alpha = .95$. For the other graphs this feature is not visible for the curve of trace-distance. 
			
	\section{Conclusion}
		
		The problem of ranking the vertices in graphs is of central importance in network analysis. Over the years, many proposals for ranking algorithms have emerged. A number of them are extremely well-known and widely applied; for instance, Google's PageRank. The recent surge of interest in quantum computing has encouraged several authors to explore the role of quantum walks in the design of ranking algorithms. This is an interesting concept which deserves further investigations. The present work belongs to this research field, which is currently active. It focuses in particular on the use of open quantum walks for node ranking in undirected and directed graphs.
		
		This article has two contributions. Firstly, it generalizes the idea of the discrete-time open quantum walk and makes it applicable to all directed and undirected graphs. Earlier, this quantum walk was defined for a limited class of graphs. Secondly, it discusses a new qPageRank for all graphs. 
		
		Corresponding to every directed edge in the graph, we construct a Kraus operator using Weyl unitary matrices. Therefore, the walker state is transported through different quantum channels for different vertices in the network. In contrast, every vertex applies the same quantum channel on the state in the standard definition of discrete time open quantum walk. Therefore, our quantum walk is a generalization of the existing idea in literature. 
		
		We apply the new idea of discrete-time open quantum walk for ranking the vertices in a graph. We produce a new quantum PageRank algorithm using this quantum walk. To define the qPageRank we maintain the procedure to calculate the classical PageRank in a quantum mechanical way. As our definition of the quantum walk fits with all graphs, the qPageRank can also be calculated for all graphs. We have considered different types of graphs for our numerical experiments. We observe that the new quantum ranking matches with classical PageRank exactly for the undirected graphs.
		
		The idea of an open quantum walk can be utilized to model the interactions between the environment and the quantum walker \cite{rani2024non}. The other models of the quantum walk are unable to incorporate the interaction with the environment. Therefore, the open quantum walk based model of qPageRank is more advantageous than the other proposals of qPageRank. An interested reader may investigate the effect of quantum noise on qPageRank procedures.
	
	\section*{Acknowledgment}
		The author is thankful to Monika Rani and Subashish Banerjee for some discussions. Also, the author like to thank anonymous reviewers for their comments to improve the article.

	\section*{Funding}
		This work was supported by the SERB Funded Project entitled “Transmission of quantum information using perfect state transfer” (Grant no. CRG/2021/001834).

	\section*{Data Availability Statement}
	
		Python programs in form of notebooks related to this article are available in the \href{https://github.com/}{GitHub} repository \url{https://github.com/dosupriyo/qPageRank}.

%	\bibliographystyle{unsrt}
%	\bibliography{library}

\end{document}